\documentclass[amsmath,amssymb,aps,prd,two column,nofootinbib]{revtex4-2}
\usepackage{graphicx}
\usepackage{subcaption} 
\usepackage{mathtools}
\usepackage{url}
\usepackage[normalem]{ulem}
\usepackage{caption}
\usepackage{enumitem}
\usepackage{float}
\usepackage{cancel}

\usepackage[colorlinks=true, linkcolor=blue, citecolor=red, urlcolor=red]{hyperref}

\newcommand{\nn}{\nonumber}

\newcommand{\ep}{\epsilon}
\newcommand{\om}{\omega} 
\newcommand{\Om}{\Omega}

\newcommand{\del}{\partial}

\newcommand{\la}{\lambda}
\newcommand{\al}{\alpha}
\newcommand{\bt}{\beta}
\newcommand{\Ga}{\Gamma}
\newcommand{\ga}{\gamma}

\newcommand{\orcid}[1]{\href{https://orcid.org/#1}{\includegraphics[width=8pt]
{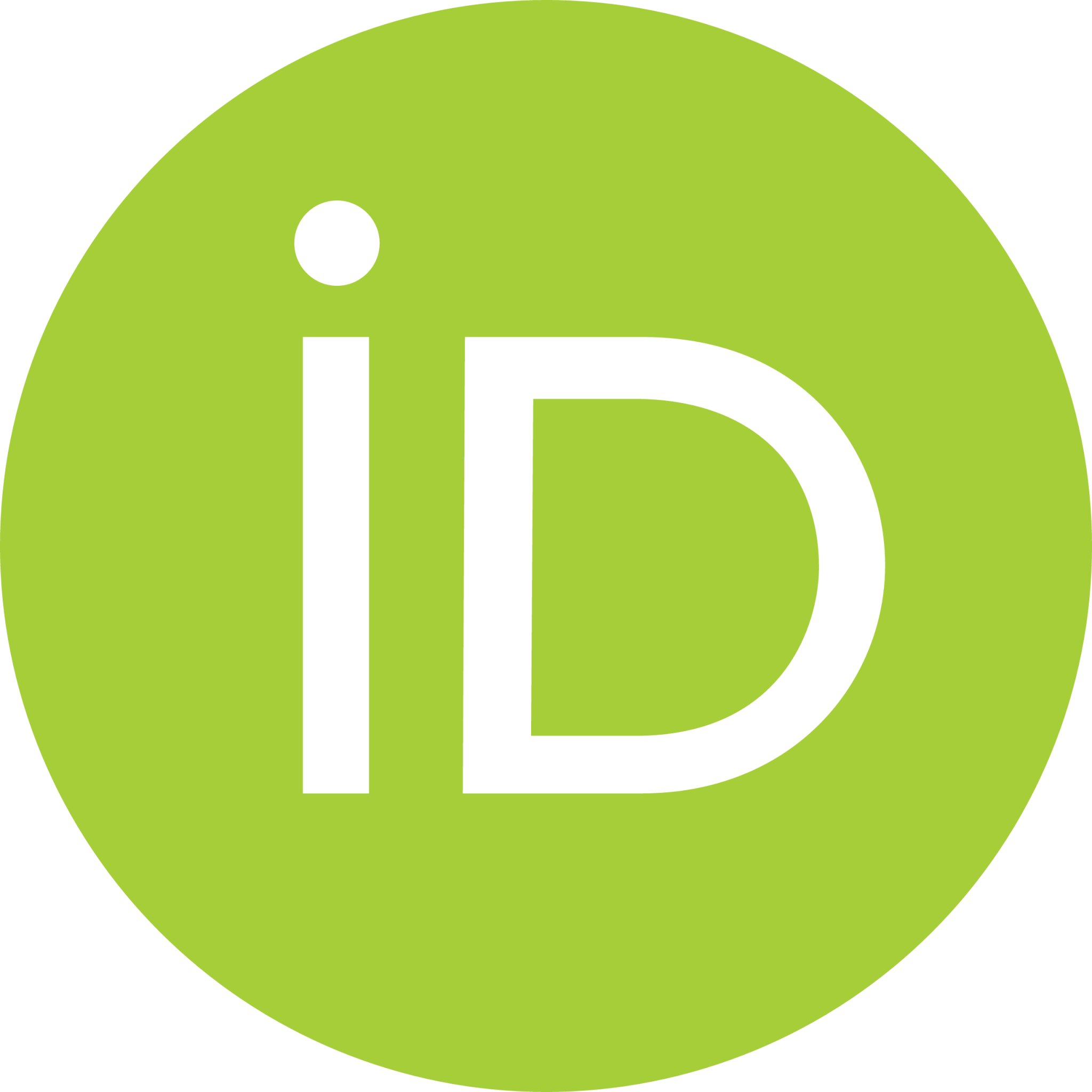}}}

\begin{document}
\title{Shear Viscosity and Electrical Conductivity of Rotating Nuclear Medium in Hadron Resonance Gas and Nambu-Jona Lasinio Models}
\author{Ashutosh Dwibedi\orcid{0009-0004-1568-2806}$^1$, Dani Rose J Marattukalam\orcid{0009-0006-8204-8148}$^{1}$, Nandita Padhan\orcid{0009-0008-6857-2650}$^2$,
Dushmanta Sahu\orcid{0000-0001-8980-1362}$^{3}$, 
Jayanta Dey\orcid{0000-0002-0894-6402}$^{1,4}$, 
Kangkan Goswami\orcid{0000-0002-0476-1005}$^{5}$, Arghya Chatterjee\orcid{0000-0002-2895-6085}$^2$, 
Sabyasachi Ghosh\orcid{0000-0003-1212-824X}$^1$, and Raghunath Sahoo\orcid{0000-0003-3334-0661}$^{5}$	
}
\affiliation{$^1$Department of Physics, Indian Institute of Technology Bhilai, Kutelabhata, Durg, 491002, Chhattisgarh, India}
\affiliation{$^{2}$Department of Physics, National Institute of Technology Durgapur, Durgapur, 713209, West Bengal, India}
\affiliation{$^{3}$Instituto de Ciencias Nucleares, Universidad Nacional Autónoma de México, Apartado Postal 70-543,
	México Distrito Federal 04510, México}
\affiliation{$^{4}$Bogoliubov Laboratory of Theoretical Physics, Joint Institute for Nuclear Research, Dubna, 141980, Russia,}
\affiliation{$^{5}$Department of Physics, Indian Institute of Technology Indore, Simrol, Indore 453552, India}

\begin{abstract}

Motivated by recent observations of spin polarization and alignment in heavy-ion collisions, we study the impact of rotation on the transport properties of strongly interacting matter within kinetic theory in the relaxation time approximation. Our analysis focuses on the anisotropic shear viscosity—parallel ($\eta_{\parallel}$), perpendicular ($\eta_{\perp}$), and Hall ($\eta_{\times}$)—and electrical conductivity—$\sigma_{\parallel}$, $\sigma_{\perp}$, and $\sigma_{\times}$—induced by the Coriolis force in a rotating medium. We employ two approaches: a combined quark–gluon plasma–hadron resonance gas (QGP–HRG) framework and a two-flavor Nambu–Jona-Lasinio (NJL) model. In the QGP–HRG description, noninteracting HRG (massless partonic) degrees of freedom are used below (above) the transition temperature. In the NJL model, rotation enters through spinorial connections in the Lagrangian, and the constituent quark masses are obtained over the full temperature range. Rotation suppresses the chiral condensate and slightly enhances the transport coefficients for phenomenologically relevant angular velocities. Assuming a temperature-dependent angular velocity consistent with standard cooling, we find that $\eta_{||,\perp,\times}/s$ and $\sigma_{\perp,\times}/T$ exhibit a valley-like temperature dependence, with reduced magnitudes compared to the isotropic $\eta/s$ and $\sigma/T$ obtained without rotation. At zero net baryon density, rotation generates a sizable nondissipative Hall-like conductivity, unlike the case with magnetic fields where baryon and antibaryon contributions cancel.
\end{abstract}

\maketitle

\section{Introduction}
The primary goal of the heavy-ion collision (HIC) experiments performed in the Large Hadron Collider (LHC) and Relativistic Heavy Ion Collider (RHIC) is to study the properties of highly dense and hot quantum chromodynamics (QCD) matter~\cite{Shuryak:2003xe,Heinz:2004qz}. A plethora of theoretical and experimental studies have supported the formation of quark-gluon plasma (QGP) in the initial stage of the HIC, followed by a hadron gas phase. Many experimental observables, such as jet quenching~\cite{Casalderrey-Solana:2007knd}, collective flow~\cite{Heinz:2013th}, nuclear suppression factor ($R_{\rm AA}$) of heavy mesons~\cite{Rapp:2018qla}, etc., show signs of the initial formation of QGP in HICs. It is worth mentioning that while the above-mentioned experimental observables have been well studied for nuclear matter produced in HICs devoid of external electromagnetic fields and/or vorticity, their studies including these latter effects have been less explored. The initial colliding nuclei of an off-central HIC can have a large orbital angular momentum (OAM), a fraction of which gets transferred to the quark-gluon medium~\cite{STAR:2017ckg,Liang:2004ph,Becattini:2007sr}. Moreover, colliding nuclei, being electrically charged and moving with ultra-relativistic velocity, can create a huge transient magnetic field in the reaction zone. The dynamics of the QCD matter in the presence of magnetic fields and angular momentum open up intense theoretical research in the field of HIC. To incorporate the effect of initial OAM in medium dynamics, two different treatments exist---in one approach, the OAM is taken to be stored in the medium locally in terms of fluid vorticity~\cite{Karpenko2021}, whereas in the other approach, a globally rotating medium is considered by defining a transformation which links the inertial frame coordinates with corotating frame coordinates~\cite{Chen2021}. In the global rotation approach, the magnitude of angular velocity is essentially taken as the spatial average of the local vortices in the fluid~\cite{Fukushima:2018grm}. There exists a list of seminal papers in which quantum field theory has been explored from a globally rotating frame~\cite{Letaw1980,Vilenkin1980,Iyer:1982ah,Duffy:2002ss,Ambrus:2014uqa,Ambrus:2015lfr}. Various other studies have considered different effective field-theoretical approaches on globally rotating QCD matter to investigate associated diverse rotational effects~\cite{Chen:2015hfc,Mameda:2015ria,Jiang:2016wvv,Ebihara:2016fwa,Chernodub_2017,Chernodub:2017ref,Chernodub:2017mvp,Chernodub:2020qah,Wang:2018sur,WeiMingHua:2020eee,Sun:2021hxo,Xu:2022hql,Sun:2023kuu,Fujimoto:2021xix}.

On the other hand, on the experimental side, there is enormous interest in looking for evidence of vorticity in the medium created in HICs. This vorticity manifests itself through the spin-orbit coupling, which can generate polarization or spin alignment along the direction of the vorticity in the local fluid cell. Averaged over the entire system, this polarization points along the direction of the angular momentum of the collision~\cite{Liang:2004ph, Becattini:2007sr}. A few years ago, the STAR collaboration made a precise measurement of the average polarization of $\Lambda$ and $\Bar{\Lambda}$ hyperons in mid-central collisions (20–50 \% centrality)~\cite{STAR:2017ckg}. Since the polarization of  $\Lambda$ and $\Bar{\Lambda}$ is solely carried by the strange quark, these measurements provided a direct probe of the rotational properties of the medium. The results revealed a positive, nonzero value for the polarization vector, offering compelling evidence for the existence of strong vorticity. Using the hydrodynamics relation, the measured polarization values across $\sqrt{s_{\rm NN}}$ = 7.7 to 200 GeV correspond to average rotational vorticity of approximately $(9 \pm 1 ) \times 10^{21}$ per second~\cite{Becattini:2016gvu, STAR:2017ckg}. Furthermore, as a function of collision energy, the polarization value decreases. This trend is also consistent with the hyperon global polarization measurement done by the ALICE collaboration for Pb-Pb collisions at $\sqrt{s_{\rm NN}}=$ 2.76 and 5.02 TeV~\cite{ALICE:2019onw}. Additionally, the spin alignment of vector mesons like $\phi$ and $K^{*0}$ in HICs is linked to system vorticity, which may arise due to the large initial angular momentum of the system. Deviations in the spin density matrix element $\rho_{00}$ from 1/3 indicate finite global spin alignment, and recent experimental results from ALICE and STAR suggest that spin alignment can serve as a probe of vorticity of the medium~\cite{ALICE:2019aid, STAR:2022fan}.

Apart from the spin polarization, which emerged as a necessary observable in recent times, the momentum and its affiliated transport coefficients, along with the equation of state (EoS) of the matter produced in HIC, play a vital role in the hydrodynamic evolution of the system. The transport coefficients of both partonic and hadronic matter in the absence of magnetic fields have been explored in Refs.~\cite{Kadam:2014cua,Gorenstein:2007mw,Itakura:2007mx,Fernandez-Fraile:2009eug,Plumari:2012ep,Lang:2012tt,Noronha-Hostler:2008kkf,Puglisi:2014pda,Greif:2014oia,Ozvenchuk:2014rpa,He:2011yi,Tolos:2013kva,Banerjee:2011ra,Goswami:2023hdl,Torres-Rincon:2021yga,Shi:2018izg,Soloveva:2020hpr,Marty:2013ita,Ghosh:2020wqx,Bernhard:2019bmu}. In the presence of magnetic fields, the rotational symmetry of the medium breaks with the magnetic field vector $\vec{B}$, singling out a particular direction in space; as a result, the transport coefficients become anisotropic~\cite{Ghosh:2019ubc,Dey:2020awu,Kalikotay:2020snc,Dey:2021fbo,Satapathy:2021cjp,Das:2019wjg,Das:2019ppb,Chatterjee:2019nld,Hattori:2016cnt,Hattori:2016lqx,Satapathy:2021wex}. Same is true in case of a rotating medium, where the direction of angular velocity $\vec{\Omega}$ breaks the isotropy of the space. Recently, the anisotropic electrical conductivities~\cite{Dwibedi:2023akm} and shear viscosities~\cite{Aung:2023pjf} of a rotating gas in a non-relativistic framework and the electrical conductivity of a rotating system of hadrons~\cite{Padhan:2024edf} in a relativistic framework have been analyzed. Moreover, the heavy quarks produced during the hard collision process may undergo an anisotropic spatial diffusion while traveling through the rotating medium, as pointed out in Ref.~\cite{Dwibedi_2025}. In this paper, by extending the previous work of Ref.~\cite{Aung:2023pjf,Dwibedi:2023akm}, we have calculated the shear viscosity and electrical conductivity of the rotating nuclear matter in both the quark and hadron phases. To fulfill this purpose, we use the Boltzmann transport equation (BTE) in the rotating frame with the collision kernel replaced by the relaxation time approximation (RTA). The rotating background has been incorporated through the rotating frame metric obtained by connecting the inertial coordinates with the corotating coordinates. In this way, the apparent forces appear in BTE in the form of connection coefficients, which in turn are expressed as the derivatives of the rotating frame metric. To keep the analysis simple, we consider the terms that are linear in the angular velocity in the BTE, which amounts to ignoring the centrifugal effects while keeping the effect of the Coriolis force.

We analyze the transport coefficients in two different frameworks: a combined massless QGP-HRG framework and an NJL framework. In the combined QGP-HRG framework, the QGP is modeled with a massless gas of quarks and gluons with constant relaxation time, whereas the hadron gas phase is modeled with the popular hadron resonance gas (HRG) model with a hard sphere scattering type interaction. The ideal HRG model, consisting of a non-interacting set of identified hadrons and their resonances, holds a special place in the analysis of HIC experimental data~\cite{KARSCH2011136,GARG2013691,PhysRevC.92.054901,PhysRevC.101.035205,PhysRevC.94.014905,PhysRevC.90.024915}. The chemical freeze-out parameters are usually determined by comparing the experimental particle yields with yields of the thermalized HRG model. Moreover, the HRG model is highly successful in reproducing the thermodynamics of QCD matter below the critical point; therefore, it is also expected to give a realistic estimation of the transport coefficients at lower temperatures. The NJL model~\cite{Nambu:1961fr,Nambu:1961tp,Hatsuda:1994pi,Klevansky:1992qe}, on the other hand, is an effective field theoretical model that encapsulates important aspects of quantum chromodynamics, specifically spontaneous chiral symmetry breaking and dynamical mass creation. The NJL model allows the study of quark condensates and their role in producing constituent quark masses by substituting an effective four-fermion contact interaction for gluonic interactions. Since the NJL framework has been widely used to study the QCD phase diagram, including phase transitions at finite temperature and density, it is a useful tool for examining the characteristics of quark matter under extreme circumstances, like those present in compact astrophysical objects and in HICs. In this work, we use the 2-flavor NJL model under rotation~\cite{Jiang:2016wvv}. In the NJL framework, we also choose the same type of relaxation time model for quarks, which is chosen for the QGP-HRG analysis, i.e., a constant relaxation time above critical temperature, whereas a hard sphere scattering type of interaction below it. It is worth pointing out that the recent lattice QCD studies~\cite{Yang:2023vsw, Braguta:2024zpi} found contradictory results from those of effective models. These first-principle studies show that the chiral condensate is enhanced and the Polyakov loop is reduced due to real rotation. In other words, catalysis of condensate and confinement occur due to real rotation both in pure SU(3) gluon dynamics and staggered fermion lattice QCD. Very recently, the NJL model-based study~\cite{Jiang2022} by Yin Jiang shows that the $\Omega$ dependent running coupling $G(\Omega)$ reverses the trend of condensate to justify effective gluon coupling. A recent work~\cite{Nunes:2024hzy} considered the $T$ and $\Omega$ dependent coupling constant ($G(T, \Omega)$) constrained by the pseudo-critical temperature obtained in lattice QCD simulation at finite rotation. However, in the current work, we have studied the transport properties of rotating quark matter for the first time using the NJL model, and we have considered a fixed coupling constant for the same.

The article is organized as follows. In Sec.~\ref{sec:Formalism}, we outline the kinetic-theory formalism and derive the expressions for the anisotropic components of shear viscosity and electrical conductivity in a rotating QCD medium. The combined QGP–HRG framework and the corresponding expressions for the transport coefficients are presented in Secs.~\ref{HRGS}. In Sec.~\ref{NJLS}, we introduce the NJL model in a rotating frame and derive the grand potential and gap equation for the constituent quark mass, along with the resulting expressions for shear viscosity and electrical conductivity. Sec.~\ref{Sec:Results_Discussion} contains our results and discussion, with emphasis on the temperature and angular-velocity dependence of the transport coefficients in both hadronic and partonic regimes. Finally, we summarize our findings in Sec.~\ref{Sec:Summary}.
\section{Formalism}
\label{sec:Formalism}
\begin{figure}[H]
	\centering
	\includegraphics[width=\columnwidth]{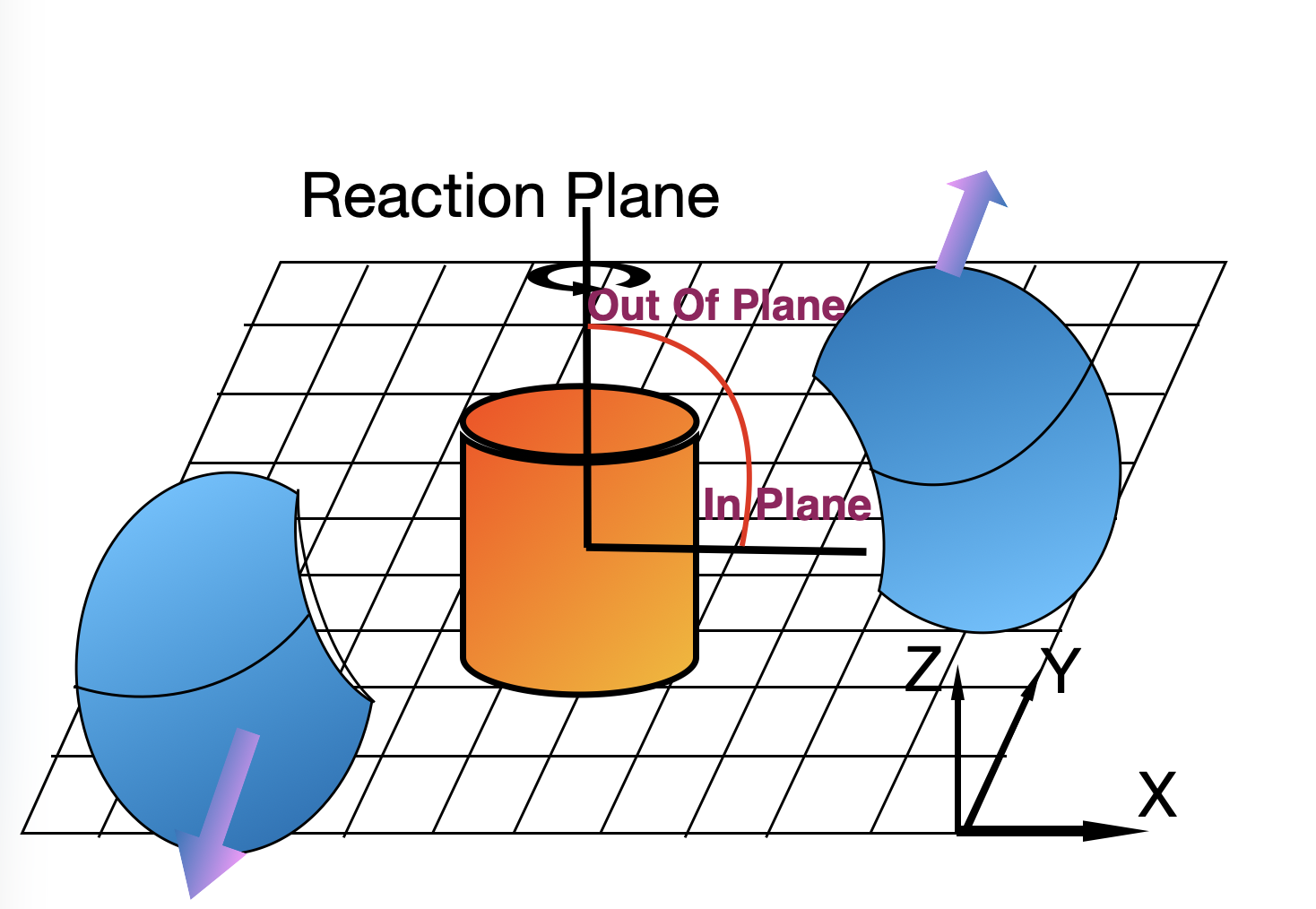}
	\caption{Schematic representation of an off-central HIC with OAM along the z-axis.}
	\label{fig:2a}
\end{figure}

\begin{figure}[H]
	\centering
	\includegraphics[width=\columnwidth]{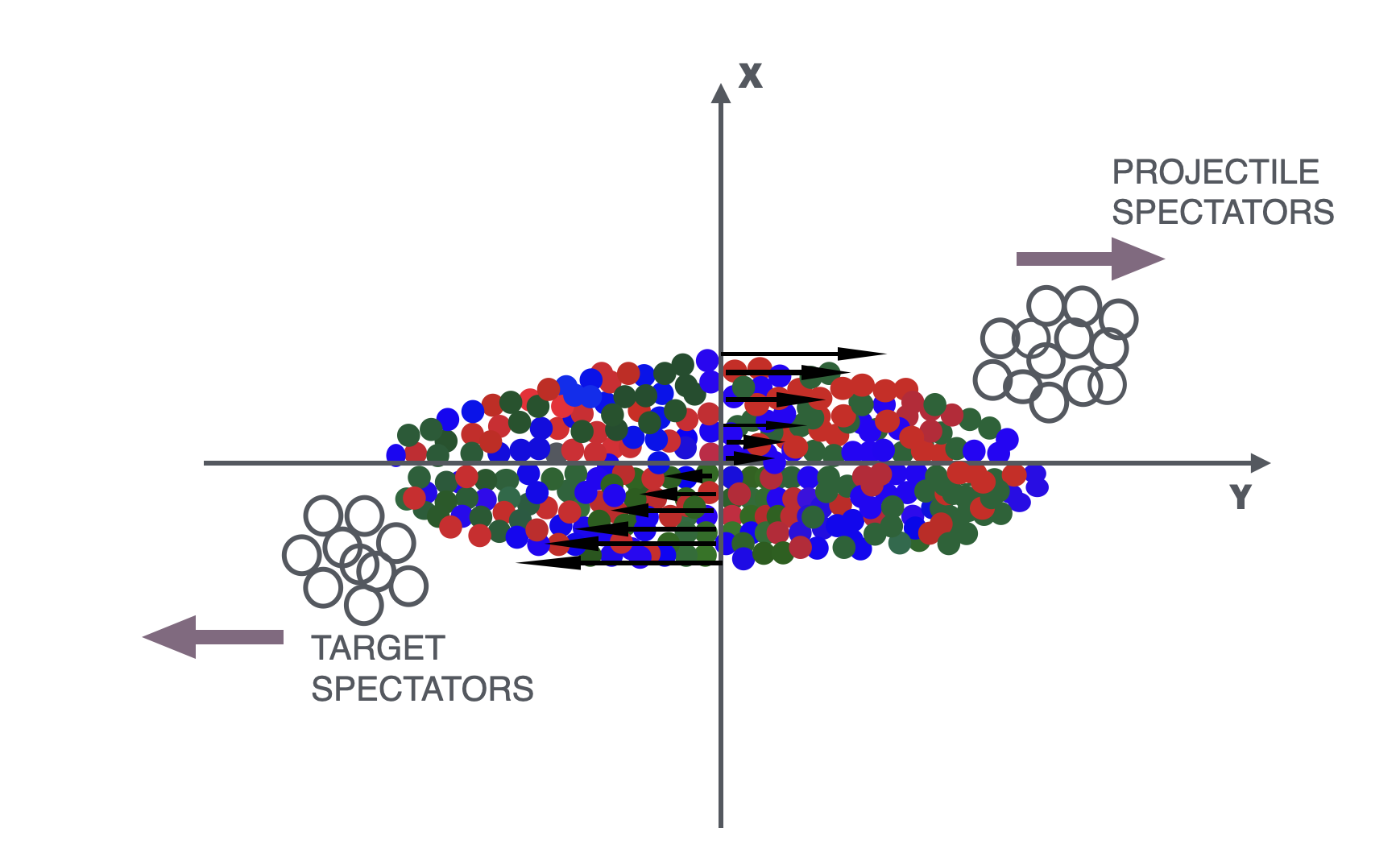}
	\caption{Illustration of rotational patterns of partons in the reaction plane during off-central HIC.}
	\label{fig:2b}
\end{figure}

We start this section with a brief introduction of the rotating kinetic theory model developed for the nuclear matter produced in off-central HIC within the non-relativistic setting in Refs.~\cite{Aung:2023pjf,Dwibedi:2023akm}, which was further generalized to the relativistic setting in Ref.~\cite{Padhan:2024edf}. By taking advantage of the fact that produced nuclear matter in off-central HIC can have a large initial OAM (see Figs.~\eqref{fig:2a} and \eqref{fig:2b}), we assume that the velocity of the medium particles has two effective parts: (1) a globally rotating part and (2) a random part on top of the globally rotating part. This effective breaking of kinematic degrees of freedom for the matter helps one to write down the BTE in the globally rotating frame and solve for the distribution function $f$ as a function of co-rotating space-time coordinates and momentum in the rotating frame.

While generalizing the non-relativistic approach of~\cite{Aung:2023pjf,Dwibedi:2023akm} to the relativistic scenario, one has to generalize the concept of non-relativistic forces in the rotating frame to a relativistic setting. For illustration, in a non-relativistic picture, the Coriolis force in a rotating frame takes the form $2m(\vec{v}\times \vec{\Omega})$, where $m$ and $\vec{v}$ are the mass and velocity of the particles. Nevertheless, the non-relativistic approximation suffers from the shortcoming that at high energies (energy scale of HICs) the produced particle moves with ultra-relativistic speed and the relation $\vec{p}=m\vec{v}$ and dispersion $E=\frac{\vec{p}^{2}}{2m}$ are no longer valid. In the realm of relativity one should expect the following changes in the Coriolis force: $2m(\vec{v}\times \vec{\Omega})\rightarrow 2\gamma_{v} m(\vec{v}\times \vec{\Omega})=2(\vec{p}\times \vec{\Omega})$\footnote{Note that the modified Coriolis force can act on massless particles moving with finite momentum.}, where $\gamma_{v}$ is the Lorentz gamma factor and can be different from the usual factor that appears in special theory of relativity.  Furthermore, the metric in the co-rotating coordinates differs from the usual Minkowski metric $\eta_{\mu\nu}$ which one uses in the special theory of relativity. All these facts should be incorporated to get the rotating frame transport equation in a mathematically legitimate way. In the following, we use the mathematical machinery of the general theory of relativity to serve the purpose. The structure of BTE in general coordinates or in the presence of gravity depends on the metric tensor $g_{\mu\nu}$ and the connection coefficients $\Gamma^{\al}_{\bt\ga}$. The co-rotating frame metric $g_{\mu\nu}$ can be readily obtained from the rotating coordinate transformation about $z-$axis as~\cite{Chernodub:2017ref,Chernodub:2020qah,Chernodub:2017mvp,Kapusta:2019sad},
\begin{eqnarray}
	&& g_{\mu\nu}=
	\begin{pmatrix}
		1-\Om^2x^{2}-\Om^2y^{2}  & \Om y & -\Om x & 0\\
		\Om y                           &      -1        &         0       & 0\\
		-\Om x                          &       0        &        -1       & 0 \\
		0                                        &       0        &         0       & -1 
		\label{S1}
	\end{pmatrix}.
\end{eqnarray}
The space-dependent metric $g_{\mu\nu}$ essentially captures non-trivial space-time geometry in the rotating frame. Its derivative can be used to get the connection coefficients in the rotating frame as~\cite{misner2017gravitation,schutz2009first,Cercignani200211}:
\begin{eqnarray}
	&&\Gamma_{\mu \lambda}^{\alpha}=\frac{1}{2}g^{\alpha \nu}\left(\frac{\del g_{\nu \mu}}{\del x^{\lambda}} +\frac{\del g_{\lambda \nu}}{\del x^{\mu}} - \frac{\del g_{\mu \lambda}}{\del x^{\nu}}\right)~.\label{S2}
\end{eqnarray}
The only non-zero connection coefficients in our case are: $\Gamma_{00}^{1}=-\Om^{2}x, \Gamma_{00}^{2}=-\Om^{2}y, \Gamma_{20}^{1}=\Gamma_{02}^{1}=-\Om, \Gamma_{10}^{2}=\Gamma_{01}^{2}=\Om$. Due to the nature of the metric $g_{\mu\nu}$, which significantly differs from $\eta_{\mu\nu}$, the covariant and contravariant components of the same vector will also differ significantly.  The four momenta for the particles are expressed as:
$p^{\al}=(\ga_{v} m,\ga_{v} m \vec{v})=(\ga_{v} m,~\vec{p})$, where~\cite{Cercignani200211,Kremer:2012nk},
\begin{eqnarray}
	&&\gamma_{v}\equiv\frac{dt}{d\tau}=\frac{1}{\sqrt{g_{00}(1+\frac{g_{0i}v^{i}}{g_{00}})^{2}-v^{2}}}~,\label{S3}
\end{eqnarray}
with the following definitions used $v^{i}\equiv \frac{dx^{i}}{dt}$ and $v^{2}\equiv(\frac{g_{0i}g_{0j}}{g_{00}}-g_{ij})v^{i}v^{j}$. Similarly, one can easily show,
\begin{eqnarray}
	p_{0}\equiv g_{0\mu}p^{\mu}=E=\sqrt{m^{2}g_{00}+(g_{0i}g_{0j}-g_{00}g_{ij})p^{i}p^{j}}~.\label{S4}
\end{eqnarray}
The covariant BTE for the rotating medium reads as~\cite{Cercignani200212,DEBBASCH20091079,DEBBASCH20091818}:
\begin{eqnarray}
	&&p^{\mu}_{r}\frac{\del f_{r}}{\partial x^{\mu}}- \Ga_{\mu \lambda}^{\al} p^{\mu}_{r}p^{\la}_{r} \frac{\partial f_{r}}{\partial p^{\al}_{r}}+m_{r} F^{\beta}_{r}\frac{\partial f_{r}}{\partial p^{\beta}_{r}}= C[f_{r}]~,\label{S5}
\end{eqnarray}
where $F^{\beta}_{r}$ is the four force and $C[f_{r}]$ is the collision kernel which arises due to the random collisions between the medium constituents.
In the presence of external electromagnetic fields, the four force is given by $m_{r}F^{\beta}_{r}=q_{r} F^{\beta\al}p_{r\al}$, where $F^{\beta\al}$ is the Faraday tensor.
Here, we are interested in calculating the transport coefficients for the rotating nuclear matter in the presence of an external electric field $\tilde{E}^{\mu}$.  Therefore, we will have $m_{r}F^{\beta}_{r}=q_{r}(\tilde{E}^{\beta} u^{\al}-\tilde{E}^{\al}u^{\beta})~p_{r\al}$ in Eq.~\eqref{S5}. The usual assumption of solving the Boltzmann kinetic equation can then be employed to split the total distribution $f_{r}$ into local equilibrium distribution $f^{0}_{r}$ and a perturbation $\delta f_{r}$, i.e., $f_{r}=f^{0}_{r}+\delta f_{r}$. The local equilibrium distributions are given by, $f^{0}_{r}=1/[e^{(p^{\al}_{r}u^{\bt}g_{\al\bt}-\mu_{r})/T}-\xi]=1/[e^{(p^{\al}_{r}u_{\al}-\mu_{r})/T}-\xi]$, where $\xi=-1$ for baryons and $\xi=+1$ for mesons. The four-vector $u^{\mu}$, and the scalars $\mu_{r}$ and $T$ occurring in the equilibrium distributions $f^{0}_{r}$ are identified with the fluid four-velocity, the chemical potential of the $r^{\text{th}}$ particle species, and temperature.  The perturbative correction $\delta f_{r}$ to the local equilibrium distribution is assumed to be small, and it contains the thermodynamic forces that drive dissipative flows.  As an application of the kinetic theory in the rotating frame, we will now proceed to derive the shear viscosity and electrical conductivity components of the rotating matter produced in HIC.

For a rotating system, the microscopic and macroscopic expressions for viscous flow or viscous stress tensor $\tau^{ij}$ (this should not be confused with the average collision time $\tau_{c}$ between particles) and electric current density $J^{i}$ can be written as,
\begin{eqnarray}
	&&\tau^{ij}=\sum_{r} \tau^{ij}_{r}=\sum\limits_{r} g_{r}\int\frac{d^{3}\vec{p}_{r}} {{(2\pi})^{3}} \frac{p^{i}_{r}p^{j}_{r}}{E_r} \delta{f_{r}}~,\label{S6}\\
	&&\tau^{i j}=-\sum_{r}\eta^{(r)ijkl}U^{kl}\equiv -\eta^{ijkl}  U^{kl}~,\label{S7}\\
	&& J^{i} =\sum_{r}J_{r}^{i}= \sum_{r} g_{r} q_{r} \int{\frac{d^{3}\vec{p}_{r}}{{(2\pi})^{3}}} \frac{p^{i}_{r}}{E_r} \delta{f_{r}},\label{new1}\\
	&& J^{i}=\sum_{r}\sigma^{(r)ij}\tilde{E}^{j}\equiv\sigma^{ij}\tilde{E}^{j},\label{new2}
\end{eqnarray}
where $U^{kl} \equiv \frac{1}{2} (\frac{\partial u^k}{\partial x^l}+\frac{\partial u^l}{\partial x^k})$ is the fluid velocity gradient, $\tilde{E}^{j}$ is the external electric field, $\eta^{ijkl}$ is the viscosity tensor and $\sigma^{ij}$ is the conductivity tensor. The macroscopic expression of viscous stress tensor $\tau^{ij}$ provided in Eq.~\eqref{S7} is reminiscent of Newton's law of viscosity.  This macroscopic expression can be compared with the microscopic kinetic theory expression provided in Eq.~\eqref{S6} for the determination of viscosity.  Similarly, the electrical conductivity can be determined by comparing the macroscopic Ohm's law given in Eq.~\eqref{new2} with the kinetic theory expression~\eqref{new1}.  For the kinetic evaluation of the viscous and conductivity tensors, we resort to the BTE in RTA.  For a system of rotating particles, we can write the BTE in RTA as \cite{Romatschke:2011qp,Cercignani200212,DEBBASCH20091079,DEBBASCH20091818}:
\begin{eqnarray}
	&&p^{\mu}_{r}\frac{\del f_{r}}{\partial x^{\mu}}- \Ga_{\mu \lambda}^{\al} p^{\mu}_{r}p^{\la}_{r} \frac{\partial f_{r}}{\partial p^{\al}_{r}}+q_{r}(\tilde{E}^{\beta} u^{\al}-\tilde{E}^{\al}u^{\beta})~p_{r\al}\frac{\partial f_{r}}{\partial p^{\beta}_{r}}\nn\\
	&&=-(u^{\al}p_{r\al})\frac{f_{r}-f^{0}_{r}}{\tau_{c}}~,\label{S8}
\end{eqnarray} 
where $\tau_{c}$ is the average collision time between the constituents.  In Eq.~\eqref{S8}, the information of the rotating space-time background has been encoded in the connection coefficients $\Ga_{\mu \lambda}^{\al}$. The second term of the LHS of Eq.~\eqref{S8} contains all the possible pseudo forces that can affect the transport properties of the system. Substituting the total distribution $f_{r}$, in Eq.~\eqref{S8}, we get the following linearized BTE, 
\begin{eqnarray}
	&&-f^{0}(1+\xi f^{0})\Bigg[\frac{p^{\mu}p^{\al}}{T} D_{\mu}u_{\al} + p^{\mu}(u_{\al}p^{\al})\del_{\mu}\frac{1}{T}-p^{\mu}\del_{\mu}\frac{\mu}{T}\nn\\
	&&	-\frac{q\tilde{E}_{\nu}p^{\nu}}{T}\Bigg]
	-\Gamma^{\sigma}_{\mu\la}p^{\mu}p^{\la}\frac{\del \delta f}{\del p^{\sigma}}=-(u^{\al}p_{\al})\frac{\delta f}{\tau_{c}}~,\label{S9}
\end{eqnarray}
where we suppressed the index $r$, which will be retained during the calculation of shear viscosity and electrical conductivity. The covariant derivative $D_{\mu}$ of fluid velocity $u_{\al}$ is defined as $D_{\mu}u_{\al}\equiv \del_{\mu}u_{\al}-\Gamma^{\sigma}_{\mu\al}u_{\sigma}$. In principle, Eq.~\eqref{S9} can be solved to obtain all the possible transport coefficients of the rotating nuclear matter.  Nevertheless, as pointed out in Ref.~\cite{Padhan:2024edf}, the calculation involved becomes cumbersome because of the space-time dependence of the rotating metric $g^{\mu\nu}$.  Therefore, in the present paper, we solve Eq.~\eqref{S9} with the same approximation that has been used in Ref.~\cite{Padhan:2024edf}.  The approximation involves ignoring the second or higher powers of $\Om x$, $\Om y$, and $\frac{\Om}{T}$, which is justified when one restricts oneself in a region that is closer to the axis of cylinder (or, away from the boundary of the causal cylinder, which is defined as the locus points satisfying $\Om \sqrt{x^{2}+y^{2}}=1$) and angular velocity $\Om$ is less than thermal energy scale ($\sim T$). In the static limit $u^{\mu}=(\frac{1}{\sqrt{g_{00}}},0)$ we have  $\Gamma^{\sigma}_{\mu\al}u_{\sigma}=0$ and $\frac{p^{\mu}p^{\al}}{T} \del_{\mu}u_{\al}=-\frac{p^{0}p^{i}}{T}\frac{\del {u^{i}}}{\del t}-\frac{p^{i}p^{j}}{T}\del_{j}u^{i}$ giving,
\begin{eqnarray}
	-f^{0}(1+\xi f^{0})&&\Bigg[-\frac{p^{0}p^{i}}{T}\frac{\del {u^{i}}}{\del t}-\frac{p^{i}p^{j}}{T}\del_{j}u^{i} + p_{0}p^{0}\frac{\del}{\del t}\frac{1}{T}\nn\\
	&&+ p_{0}p^{i}\del_{i}\frac{1}{T}-p^{0}\frac{\del}{\del t}\frac{\mu}{T}-p^{i}\del_{i}\frac{\mu}{T}+\frac{q\tilde{E}^{i}p^{i}}{T}\Bigg]\nn\\
	&& + 2 \Om p^{0}\left(p^{2}\frac{\del \delta f}{\del p^{1}}-p^{1}\frac{\del \delta f}{\del p^{2}} \right)=-p_{0}\frac{\delta f}{\tau_{c}}.\nn\\\label{S10}
\end{eqnarray}
The time derivatives of $\mu/T$, $1/T$, and $u^{i}$ occurring in Eq.~\eqref{S10} can be eliminated with the help of conservation equations~\cite{Cercignani200212}, and they finally contribute to the scalar (bulk viscous flow) and vector sector (thermal current) of the transport.  Since the present paper is planned for the calculation of the shear viscosity and electrical conductivity, we can ignore the scalar and vectorial parts of the thermodynamic fluxes occurring in the LHS of Eq.~\eqref{S10} to write,
\begin{eqnarray}
	&&\frac{f^{0}(1+\xi f^{0})}{ET}\left( p^{i}p^{j}\del_{i}u^{j}-q\tilde{E}^{i}p^{i}\right)+ 2(\vec{p}\times \vec{\Om})\cdot \frac{\del \delta f}{\del \vec{p}}\nn\\
	&&=-\frac{\delta f}{\tau_{c}}~,
	\label{S11}
\end{eqnarray}
where we assumed a rotation about the $z-$ axis, i.e., $\vec{\Om}=\Om~ \hat{k}$ (cf. Fig.~\eqref{fig:2a}). Eq.~\eqref{S11} bears a formal resemblance to the one encountered in the computation of shear viscosity and electrical conductivity in the presence of a magnetic field.  Therefore, it can be solved using a similar technique as employed in Ref.~\cite{Dey:2019axu}.  By using the definitions: $\tau_{\Om}\equiv\frac{1}{2\Om}$, $\vec{\Om}\equiv \Om \hat{\om}\implies \Om^{i}=\Om \om^{i}$, and $U^{ij}\equiv \frac{1}{2} (\frac{\del u^i}{\del x^j}+\frac{\del u^j}{\del x^i})$ we can  rewrite Eq.~\eqref{S11} as:
\begin{eqnarray}
	&&\frac{f^{0}(1+\xi f^{0})}{ET}\left(p^{i}p^{j}U^{ij} -q\tilde{E}^{i}p^{i}\right)+ \frac{1}{\tau_{\Om}} \ep^{ijk}p^{j}\om^{k} \frac{\del \delta f}{\del p^{i}}\nn\\
	&&=-\frac{\delta f}{\tau_{c}}~.
	\label{S12}
\end{eqnarray}
For the evaluation of viscous stress and conductivity tensors in Eq.~\eqref{S6} and Eq.~\eqref{new1}, one has to solve Eq.~\eqref{S12} for $\delta f$. Here, we outline the steps to solve Eq.~\eqref{S12}, delegating the detailed calculations to Appendix~\ref{ape1}.  A quick glance at Eq.~\eqref{S12}, suggests that the solution $\delta f$ should be proportional to two powers of momentum, i.e., $\delta f \propto p^{k}p^{l}$ if one has $\tilde{E}^{i}=0$ and one power of momentum, i.e., $\delta f \propto p^{k}$ if $U^{ij}=0$.
This leads to the most general guess $\delta f = \sum\limits_{n=0}^{6} C_{n}C_{n}^{kl}p^{k}p^{l}+\sum\limits_{n=0}^{2}A_{n}~A_{n}^{k}~p^{k}$, where $C_{n}$ and $A_{n}$ are functions of energy along with other possible thermodynamic variables.  The $C_{n}^{kl}$ are the seven independent velocity gradients that one can construct with the help of the tensors: $U^{ij}$, $\delta^{ij}$, $\om^{i}$, and $\ep^{ijk}$.  The $A_{n}^{k}$ are three independent vector components made up of $\tilde{E}^{i}$, $\delta^{ij}$, $\om^{i}$  and $\ep^{ijk}$. It is noteworthy that the first part of the solution $\delta f$, which involves two powers of momentum, does not contribute to the electric current density, as the corresponding integral in Eq.~\eqref{new1} vanishes. Similarly, the second part of the solution does not contribute to the viscous stress tensor, since the integral in Eq.~\eqref{S6} vanishes.  Consequently, it is sufficient to treat the two parts of the solution separately, i.e., $\delta f=\delta f^{\eta}+\delta f ^{\sigma}$, while evaluating the viscous stress tensor and the electric current density. Subsequently, we briefly address two aspects: first, the procedure for obtaining the shear viscosity coefficients, and second, the method for determining the conductivity components, assuming that the coefficients $C_{n}$ and $A_{n}$ are known.   

Out of the seven independent tensors $C_{n}^{kl}$, the first five are traceless, and the other two have non-zero traces.  The seven $C_{n}^{kl}$  can be considered as the thermodynamic forces that drive dissipative tensorial flows in the rotating medium.  The first five $C_{n}^{kl} (n=0\text{ to } 4)$ drive shear flows, and the respective proportionality factors are called shear viscosities, whereas the next two $C_{n}^{kl} (n=5,6)$ drive bulk flows and the respective proportionality factors are called bulk viscosities.  The seven velocity gradients $C_{n}^{ij}$ can be written as contraction of seven $4-$rank tensors $C_{n}^{ijkl}$ with $U^{kl}$, i.e., $C_{n}^{ij}\equiv C_{n}^{ijkl} U^{kl}$. Retaining the particle index $r$ which we suppressed while transitioning from Eq.~\eqref{S8} to Eq.~\eqref{S9}, the viscosity tensor for the $r^{\text{th}}$ particle species $\eta^{(r)ijkl}$ can be expressed in terms of the $4-$rank tensors $C_{n}^{ijkl}$ as,
\begin{eqnarray}
	\eta^{(r)ijkl}&=& \eta^{r}_0 C_{0}^{ijkl}+\eta^{r}_1C_{1}^{ijkl}+\eta^{r}_2 C_{2}^{ijkl}+\eta^{r}_3C_{3}^{ijkl}\nn\\
	&&+\eta^{r}_4C_{4}^{ijkl}+\zeta^{r}_5C_{5}^{ijkl}+\zeta^{r}_6 C_{6}^{ijkl},
	\label{S13}    
\end{eqnarray} 
where $\eta^{r}_{n}$ (for $n= 0 \text{ to }4$) and $\zeta^{r}_{n}$ (for $n= 5, 6$) are identified with shear and bulk viscosity coefficients, respectively.
Using the macroscopic version of Newton's law provided in Eq.~(\ref{S7}), we have,
\begin{eqnarray}
	\tau^{i j}&=&-\sum_{r}\eta^{(r)ijkl}U^{kl}=-\sum\limits_{n=0}^{4}\eta_{n}C_{n}^{ij}-\sum\limits_{n=5}^{6}\zeta_{n}C_{n}^{ij},\nonumber\\
	\label{S14}
\end{eqnarray}
where we defined the total shear and bulk viscosity coefficients of the rotating medium as the sum of the contributions from individual particles, i.e., $\eta_{n}=\sum_{r}\eta^{r}_{n}$ and $\zeta_{n}=\sum_{r}\zeta^{r}_{n}$. Since the present article is structured for the calculation of shear viscosity coefficients of the rotating QCD matter, we will keep only the terms that correspond to the shear stress tensor or shear flow in Eq.~\eqref{S6} and Eq.~\eqref{S14} and rewrite them as follows:
\begin{eqnarray}\
	&&\pi^{ij}=\sum\limits_{r} g_{r}\int\frac{d^{3}\vec{p}_{r}} {{(2\pi})^{3}} \frac{p^{i}_{r}p^{j}_{r}}{E_r} \delta{f_{r}}~,\label{S15}\\
	&&\pi^{ij}=-\sum\limits_{n=0}^{4}\eta_{n}C_{n}^{ij}~.\label{S16}
\end{eqnarray}
For the kinetic evaluation of the shear flow $\pi^{ij}$ we will substitute $\delta f^{\eta}_{r}=\sum\limits_{n=0}^{4}C^{r}_{n}C_{n}^{kl}p_{r}^{k}p_{r}^{l}$ in Eq.~\eqref{S15} to obtain,
\begin{eqnarray}
	\pi^{ij}&=&\sum\limits_{r} g_{r}\int\frac{d^{3}\vec{p}}{(2\pi)^3 E}\sum_{n=0}^{4}C_n C^{kl}_{n} ~p^i p^jp^kp^l\nn\\
	\implies  \pi^{ij}&=&\sum\limits_{r} g_{r}\int\frac{d^{3}p}{(2\pi)^3 E} \sum_{n=0}^{4} C_n C^{kl}_{n}\nonumber\\
	&& (\delta^{ij} \delta^{kl} +\delta^{ik}\delta^{jl} +\delta^{il} \delta^{jk})~ \frac{p^4}{15}\nn\\
	\implies  \pi^{ij} &=&\sum_{n=0}^{4} C^{ij}_n\sum\limits_{r}\frac{2g_{r}}{15}\int \frac{d^3p}{(2\pi)^3 E}C_n~p^4 ,
	\label{S17}
\end{eqnarray}
where for the notational simplicity we suppressed the particle index $r$ in the unknown coefficients $C_{n}$ and momentum components $p^{i}$. We also used the identities, $\int p^ip^jp^kp^l d^{3}\vec{p}=\int\frac{p^4}{15}(\delta^{ij}\delta^{kl}+\delta^{ik}\delta^{jl}+\delta^{il}\delta^{jk})d^3p, (d^3 p\equiv4\pi p^2dp)$ and $C^{kl}_n (\delta^{ij} \delta^{kl} +\delta^{ik} \delta^{jl} +\delta^{il}\delta^{jk})=2C^{ij}_n$. The unknown coefficients $C_{n}$ appearing in Eq.~\eqref{S17} are explicitly calculated in Appendix~\ref{ape1}. Using the expression of $C_{n}$ obtained in Eq.~\eqref{AA6} and comparing it with Eq.~\eqref{S16} we obtain the following expressions for shear viscosity components,
\begin{eqnarray}
	&&\eta_1=\sum\limits_{r}\frac{g_{r}}{15T}\frac{\tau_c}{1+4(\tau_c/\tau_\Om)^2}\int\frac{d^3p}{(2\pi)^3}\frac{p^4}{E^2}f^{0}(1+\xi f^{0})\nn\\
	&&\eta_2=\sum\limits_{r}\frac{g_{r}}{15T}\frac{\tau_c}{1+(\tau_c/\tau_\Om)^2}\int\frac{d^3p}{(2\pi)^3}\frac{p^4}{E^2}f^{0}(1+\xi f^{0})\nn\\
	&&\eta_3=\sum\limits_{r}\frac{g_{r}}{15T}\frac{2\tau_c(\tau_c/\tau_\Om)}{1+4(\tau_c/\tau_\Om)^2}\int\frac{d^3p}{(2\pi)^3}\frac{p^4}{E^2}f^{0}(1+\xi f^{0})\nn\\
	&&\eta_4=\sum\limits_{r}\frac{g_{r}}{15T}\frac{\tau_c(\tau_c/\tau_\Om)}{1+(\tau_c/\tau_\Om)^2}\int\frac{d^3p}{(2\pi)^3}\frac{p^4}{E^2}f^{0}(1+\xi f^{0})\nn\\
	\label{Shear_Rel}
\end{eqnarray}
The shear viscosity component in the absence of rotation $\eta_{0}\equiv \eta$ is given by~\cite{Dey:2019axu},
\begin{eqnarray}
	&&\eta_0=\sum\limits_{r}\frac{g_{r}}{15T} \tau_c \int\frac{d^3p}{(2\pi)^3}\frac{p^4}{E^2}f^{0}(1+\xi f^{0})~.\label{Isoshe}
\end{eqnarray}
One can define the perpendicular $\eta_{\perp}$, parallel $\eta_{||}$ and Hall $\eta_{\times}$ viscosity as~\cite{Dey:2019axu}, $\eta_{1}\equiv \eta_{\perp}$, $\eta_{2}\equiv \eta_{||}$, and $\eta_{4}\equiv \eta_{\times}$.

Three independent vectors $A_{n}^{k}$ can be considered as thermodynamic forces that drive the electric current in the rotating medium.  The vectors $A_{n}^{k}$ can be expressed as contraction of three $2-$rank tensors $A_{n}^{kl}$ with $\tilde{E}^{l}$, i.e., $A_{n}^{k}\equiv A_{n}^{kl} \tilde{E}^{l}$. Retaining the particle index $r$, the electrical conductivity tensor for the $r^{\text{th}}$ particle species $\sigma^{(r)ij}$ can be expressed  in terms of the $2-$rank tensors $A_{n}^{kl}$ as, 
\begin{eqnarray}
	\sigma^{(r)ij}&=& \sigma^{r}_0 A_{0}^{ij}+\sigma^{r}_1A_{1}^{ij}+\sigma^{r}_2 A_{2}^{ij},
	\label{new3}    
\end{eqnarray}
where $\sigma_{n}^{r}$ (for $n=0$ to $2$) are identified with the coefficients of electrical conductivity. Employing Ohm's law given in Eq.~\eqref{new2}, we have,
\begin{eqnarray}
	J^{i}=\sum_{r}\sigma^{(r)ij}\tilde{E}^{j}=\sum_{n=0}^{2}\sigma_{n}A_{n}^{i}~,
	\label{new4}    
\end{eqnarray}
where the total conductivity components are defined as, $\sigma_{n}=\sum_{r}\sigma_{n}^{r}$. For the evaluation of these conductivity components from kinetic theory we substitute $\delta f^{\sigma}_{r}=\sum\limits_{n=0}^{2}A^{r}_{n}~A_{n}^{k}~p_{r}^{k}$ in Eq.~\eqref{new1} to get,
\begin{eqnarray}
	J^{i}&=&\sum\limits_{r} g_{r}q_{r}\int\frac{d^{3}\vec{p}}{(2\pi)^3 E}\sum_{n=0}^{2}A_n A^{k}_{n} ~p^i p^k\nn\\
	\implies  J^{i}&=&\sum\limits_{r} q_{r}g_{r}\int\frac{d^{3}p}{(2\pi)^3 E} \sum_{n=0}^{2} A_n A^{k}_{n}~ \frac{p^{2}}{3} \delta^{ik}\nn\\
	\implies  J^{i} &=&\sum_{n=0}^{2} A^{i}_n\sum\limits_{r}\frac{g_{r}q_{r}}{3}\int \frac{d^3p}{(2\pi)^3 E}A_n~p^2~.
	\label{new5}
\end{eqnarray}
Once again we suppressed the particle index $r$ in the momentum and in the unknown coefficients $A_{n}$ while writing Eq.~\eqref{new5}. Substituting the value of $A_{n}$ from Eq.~\eqref{sigmasolex} in Eq.~\eqref{new5}, we get the conductivity components as
\begin{eqnarray}
	\sigma_{n}=\sum\limits_{r}\frac{g_{r} q_{r}^2}{3T} \int \frac{d^3 p}{(2\pi)^3}~\frac{p^2}{E^2}\frac{\tau_c\left(\frac{\tau_{c}}{\tau_{\Omega}}\right)^{n}}{1+\Big(\frac{\tau_c}{\tau_{\Omega}}\Big)^2} f^{0}(1+\xi f^{0}).
	\label{new6}
\end{eqnarray}
The parallel or the conductivity in the absence of rotation, perpendicular and the hall conductivities are defined as, $\sigma_{||}\equiv \sigma_{0}+\sigma_{2}$, $\sigma_{\perp}\equiv\sigma_{0}$ and $\sigma_{\times}\equiv\sigma_{1}$, respectively.

After this general discussion of the rotating-frame setup and the evaluation of the shear viscosity and electrical conductivity components in the rotating frame, we shift our attention to two different approaches for a realistic estimation of the viscosities and conductivities of the matter produced in HICs over the full temperature range.  In Sec.~\ref{HRGS}, we employ a combined description of a massless QGP and an HRG, and present the final expressions for the conductivity and viscosity within this framework. Subsequently, in Sec.~\ref{NJLS}, we present the same quantities within the ambit of the NJL model. 

\subsection{Shear viscosity and electrical conductivity for massless QGP-HRG}\label{HRGS}
The hadronic phase of matter created in HICs can be effectively described by HRG model~\cite{KARSCH2011136,GARG2013691,PhysRevC.92.054901,PhysRevC.101.035205,PhysRevC.94.014905,PhysRevC.90.024915}. This model is widely accepted in the community for the characterization of the thermodynamics~\cite{Karsch:2003zq,Braun-Munzinger:2015hba}, conserved charge fluctuations~\cite{Begun:2006jf,Nahrgang:2014fza,HotQCD:2012fhj,Bhattacharyya:2013oya,Chatterjee:2016mve}, as well as transport coefficients~\cite{Gorenstein:2007mw,Noronha-Hostler:2012ycm,Tiwari:2011km,Pradhan:2022gbm,Noronha-Hostler:2008kkf,Kadam:2014cua,Zhang:2019uct,Samanta:2017ohm,Ghosh:2019fpx,Ghosh:2018nqi,Rocha:2024rce} of the created hadron gas in HICs. The HRG model offers a grand canonical ensemble treatment of the hadron gas by including all the degrees of freedom associated with the system, i.e., hadrons and their resonances. It has been shown by the $S$-matrix calculation that in the presence of narrow resonances, the interacting gas of hadrons can be approximated by an ideal gas comprising hadrons and their resonances~\cite{Dashen:1969ep,Dashen:1974jw}. Here, we will model the hadron gas by an ideal gas of non-interacting point-like hadrons and resonances up to mass 2.6 GeV as listed in Ref.~\cite{ParticleDataGroup:2008zun}. In the grand canonical ensemble approach of the ideal HRG model, one expresses grand potential $\Phi^{\rm HRG}_{\rm G}=-P^{\rm HRG}~V$ as~\cite{Ratti2021},
\begin{eqnarray}
	\Phi^{\rm HRG}_{\rm G}&=& -T\sum_{B}\frac{Vg_{B}}{2\pi^2}\int p^2 dp~ \ln\left[1+ e^{-E_{B}/T}\right]\nn\\ &+&T\sum_{M}\frac{Vg_{M}}{2\pi^2}\int p^2 dp~ \ln\left[1- e^{-E_{M}/T}\right],\label{T3}
\end{eqnarray} 
where we consider zero chemical potential for all hadrons with two separate summations for baryons (B) and mesons (M). In Eq.~(\ref{T3}), $m_{B}$, $E_{B}=\sqrt{p^{2}+m_{B}^{2}}$, and $g_{B}=2S_{B}+1$ are equal to the mass, energy, and spin degeneracy of the baryons with spin $S_{B}$, respectively. Similarly, we have $m_{M}$, $E_{M}=\sqrt{p^{2}+m_{M}^{2}}$, $g_{M}=2S_{M}+1$ equal to the mass, energy, and spin degeneracy of the mesons with spin $S_{M}$, respectively. The entropy density can be obtained from Eq.~\eqref{T3} as,
\begin{eqnarray}
	&& s^{\rm HRG}=-\frac{1}{V}\frac{\partial \Phi^{\rm HRG}_{\rm G}}{\partial T}= \frac{\mathcal{E}^{\rm HRG}+P^{\rm HRG}}{T}, \label{T4}
\end{eqnarray}
where energy density $\mathcal{E}^{\rm HRG}$ and pressure $P^{\rm HRG}$ of the HRG are given by,
\begin{eqnarray}
	\mathcal{E}^{\rm HRG}&=&\sum_{B}  \frac{g_{B}}{2\pi^{2}} \int \frac{p^{2}dp}{e^{E_{B}/T}+ 1} E_{B}\nn\\
	&+& \sum_{M}  \frac{g_{M}}{2\pi^{2}} \int \frac{p^{2}dp}{e^{E_{M}/T}- 1} E_{M}~,\label{T5}\\
	P^{\rm HRG}&=& \sum_{B} \frac{g_{B}}{2\pi^{2}} \int \frac{p^{2}dp}{e^{E_{B}/T} +1} \left(\frac{p^{2}}{3E_{B}}\right)\nn\\
	&+&\sum_{M} \frac{g_{M}}{2\pi^{2}} \int \frac{p^{2}dp}{e^{E_{M}/T}- 1} \left(\frac{p^{2}}{3E_{M}}\right)~.\label{T6}
\end{eqnarray}
Similarly, we can express the shear viscosity components given in Eqs.~\eqref{Isoshe} and ~\eqref{Shear_Rel} by two separate summations for baryons (B) and mesons (M) as follows,  
\begin{eqnarray}
	\eta^{\rm HRG}&=& \sum_{B} \frac{g_B}{15T}\int \frac{d^{3}p}{(2\pi)^3}\tau^{B}_c \frac{p^4}{E_{B}^{2}}f_{B}^0(1-f_{B}^0)\nn\\
	&+& \sum_{M} \frac{g_M}{15T}\int \frac{d^{3}p}{(2\pi)^3}\tau^{M}_c \frac{p^4}{E_{M}^{2}}f_{M}^0(1+f_{M}^0),\label{Heta0}\\
	\eta^{\rm HRG}_{\perp,||,\times}&=&\sum_{B} \frac{g_B}{15T}\int \frac{d^{3}p}{(2\pi)^3} \tau^{B}_{\eta\{\perp,||,\times\}}~\frac{p^4}{E_{B}^{2}}f_{B}^0(1-f_{B}^0)\nn\\
	&+& \sum_{M} \frac{g_M}{15T}\int \frac{d^{3}p}{(2\pi)^3}\tau^{M}_{\eta\{\perp,||,\times\}}~ \frac{p^4}{E_{M}^{2}}f_{M}^0(1+f_{M}^0),\nn\\
	\label{Heta1}
\end{eqnarray}
where $f^{0}_{B}=1/(e^{E_{B}/T}+1)$, $f^{0}_{M}=1/(e^{E_{M}/T}-1)$, and $\tau_{c}^{B,M}$ are the relaxation times of hadrons (baryons and mesons). For conductivity one obtains~\cite{Padhan:2024edf,Padhan:2025nps},
\begin{eqnarray}
	\sigma^{\rm HRG}_{||}&=&\sum\limits_{B}\frac{g_{B} q_{B}^2}{3T} \int \frac{d^3 p}{(2\pi)^3}~\frac{p^2}{E_{B}^2} \tau_{c}^{B} f_{B}^{0}(1- f_{B}^{0})\nn\\
	&+& \sum\limits_{M}\frac{g_{M} q_{M}^2}{3T} \int \frac{d^3 p}{(2\pi)^3}~\frac{p^2}{E_{M}^2}\tau_{c}^{M} f_{M}^{0}(1+ f_{M}^{0})~,\nn\\
	\label{new7}\\
	\sigma^{\rm HRG}_{\perp,\times}&=&\sum\limits_{B}\frac{g_{B} q_{B}^2}{3T} \int \frac{d^3 p}{(2\pi)^3}~\frac{p^2}{E_{B}^2} \tau_{\sigma\{\perp,\times\}}^{B} f_{B}^{0}(1- f_{B}^{0})\nn\\
	&+& \sum\limits_{M}\frac{g_{M} q_{M}^2}{3T} \int \frac{d^3 p}{(2\pi)^3}~\frac{p^2}{E_{M}^2} \tau_{\sigma\{\perp,\times\}}^{M} f_{M}^{0}(1+ f_{M}^{0})~.\nn\\
	\label{new8}
\end{eqnarray} 
The effective relaxation times occurring in the expressions of the shear viscosity and conductivity are defined as
\begin{eqnarray}
	&& \tau^{ B,M}_{\eta\perp}=\frac{\tau^{B,M}_{c}}{1+4\big(\frac{\tau^{B,M}_c}{\tau_\Om}\big)^2}~,\nn\\
	&&\tau^{ B,M}_{\eta ||}=\frac{\tau^{B,M}_c}{1+\big(\frac{\tau^{B,M}_c}{\tau_\Om}\big)^2}=\tau^{ B,M}_{\sigma \perp}~,\nn\\
	&&\tau^{ B,M}_{\eta \times}\equiv\frac{\tau^{B,M}_c\big(\frac{\tau^{B,M}_c}{\tau_\Om}\big)}{1+\big(\frac{\tau^{B,M}_c}{\tau_\Om}\big)^2}=\tau^{ B,M}_{\sigma \times}\label{relaxH}~.
\end{eqnarray} 

We can get the shear viscosity and conductivity components for a massless rotating QGP by replacing $p$ by $E$ and summing over the quark (q) and gluon (g) degrees of freedom with appropriate degeneracies in Eqs.~(\ref{Shear_Rel}),~(\ref{Isoshe}) and \eqref{new6} as,
\begin{eqnarray}
	&&\eta^{\rm QGP}_{0}\equiv \eta ^{\rm QGP}=\frac{19 \pi^{2}}{45}\tau_{c}T^{4}~,\label{isoQ}\\
	&& \sigma^{\rm QGP}\equiv \sigma_{||}^{\rm QGP}=\frac{2e^{2}}{9}\tau_{c}T^{2}~.\label{isosiQ}
\end{eqnarray}
Similarly, one can express the perpendicular, parallel, and Hall viscosity and perpendicular and Hall conductivity as,
\begin{eqnarray}
	&&\eta^{\rm QGP}_{\perp,||,\times}=\eta^{\rm QGP}_{1,2,4}=\frac{19 \pi^{2}}{45}\tau_{\eta\{\perp,||,\times\}}T^{4}~,\label{shQ}\\
	&& \sigma_{\perp,\times}^{\rm QGP}=\sigma_{0,1}^{\rm QGP}=\frac{2e^{2}}{9}\tau_{\sigma\{\perp,\times\}}T^{2}~,\label{siQ}
\end{eqnarray}
where the effective relaxation times are defined in a manner similar to Eq.~\eqref{relaxH}. Entropy density $s^{\rm QGP}$ for the massless rotating QGP can be written as:
\begin{eqnarray}
	&& s^{\rm QGP}=\frac{19\pi^{2}}{9}T^{3}\label{Qent}~.
\end{eqnarray}

\subsection{Shear Viscosity and Electrical Conductivity for the NJL Model}\label{NJLS}
The Lagrangian for the two-flavor NJL model in a rotating frame is given by~\cite{Jiang:2016wvv,Chen:2015hfc,Chernodub_2017,Ebihara:2016fwa,Wang:2018sur,Chernodub:2017ref},
\begin{eqnarray}
	\mathcal{L} = \bar{\psi}[i\gamma^{\mu}(\partial_{\mu}+\Gamma_{\mu})-m]\psi+G_s[(\bar{\psi}\psi)^2+(\bar{\psi}i\gamma_5\vec{\sigma}\psi)^2],\nn\\
	\label{A1}
\end{eqnarray}
where the quark fields $\psi\equiv
\begin{pmatrix}
	\psi_{u} \\
	\psi_{d} \\
\end{pmatrix}$. The spinorial affine connection $\Gamma_{\mu}$ can be expressed as $\Gamma_{\mu}=\frac{1}{8}\omega_{\mu ab} [\gamma^{a},\gamma^{b}]$. The spin connections $\omega_{\mu ab}$ is written in terms of the metric $g^{\mu\nu}$, vierbein fields (tetrads) $e^{\mu}_{~a}$ and the affine connections (Christoffel symbols) $\Gamma^{\mu}_{\al\bt}$ as~\cite{Pollock:2010zz,Yepez:2011bw,Kapusta:2019sad,RevModPhys.29.465}, $\om_{\mu ab}=g_{\nu\bt}e^{\nu}_{~a}(\del_{\mu}e^{\bt}_{~b}+e^{\la}_{~b}\Gamma^{\bt}_{\mu\la})$. The Greek and Latin indices are used for the co-rotating coordinate space ($\mu=t,x,y,z$) and the inertial tangent space $(a=0,1,2,3)$ defined by the vierbein fields, respectively. The vierbein which define the local tangent space is chosen as~\cite{Kapusta:2019sad,Ebihara:2016fwa}, $e^{t}_{~0}=e^{x}_{~1}=e^{y}_{~2}=e^{z}_{~3}=1,~ e^{x}_{~0}=\Om y,~ e^{y}_{~0}=-\Om x$ and zero otherwise. 
The Lagrangian (\ref{A1}) in the mean-field approximation for isospin symmetric matter can be written as~\cite{Wang:2018sur,Chen2021},
\begin{eqnarray}
	\mathcal{L}=\bar{\psi}[i\gamma^{\mu}(\partial_{\mu}+\Gamma_{\mu})-M]\psi-\frac{(M-m)^{2}}{4G_{S}},\label{A2}
\end{eqnarray}
where the constituent quark mass $M\equiv m-2G_{S}\langle\bar{\psi}\psi \rangle$. The constituent quark mass $M$ is the dynamically generated mass that comes into play as a result of quark self-energy~\cite{Buballa:2003qv}. Starting from the mean-field approximated Lagrangian (\ref{A2}) and using the standard thermal field theory techniques; one obtains the following grand potential energy $\Phi^{\rm NJL}_{\rm G}$ for the rotating system of quarks in the local density approximation~\cite{Wang:2018sur,Chen2021},
\begin{widetext}
	\begin{eqnarray}
		\Phi^{\rm NJL}_{\rm G}&=&\int d^{3}\vec{x}\bigg[\frac{(M-m)^{2}}{4G_{S}}-\frac{N_{c}N_{f}}{8\pi^{2}}\sum_{k=-\infty}^{\infty}\int_{0}^{\Lambda} dp_{\perp}^{2} \int_{-\sqrt{\Lambda^{2}-p_{\perp}^{2}}}^{\sqrt{\Lambda^{2}-p_{\perp}^{2}}} dp_{z} ~ E \left(J^{2}_{k}(p_{\perp}\rho)+J^{2}_{k+1}(p_{\perp}\rho)\right)\nn\\
		&-&\frac{TN_{c}N_{f}}{8\pi^{2}}\sum_{k=-\infty}^{\infty}\int dp_{\perp}^{2} \int dp_{z}\left(\ln(1+e^{-\bt(E-(k+\frac{1}{2})\Om)}) +\ln (1+e^{-\bt(E+(k+\frac{1}{2})\Om)})\right)\left(J^{2}_{k}(p_{\perp}\rho)+J^{2}_{k+1}(p_{\perp}\rho)\right)\bigg],\nonumber\\
		\label{A3}
	\end{eqnarray}
	where $J_{k}$ is the $k$th order Bessel function of the first kind and $p_{\perp}\equiv \sqrt{p_{x}^{2}+p_{y}^{2}}$, $\rho\equiv\sqrt{x^{2}+y^{2}}$ are the radial momentum and coordinate respectively. The particle energy $\Tilde{E}$ in a co-rotating frame can be expressed as~\cite{Ambru__2014,Ambru__2016}, $\Tilde{E}=E-(k+\frac{1}{2})\Om$, where $E=\sqrt{p_{\perp}^{2}+p_{z}^{2}+M^{2}}$. Since the NJL model is not renormalizable, one needs to prescribe a scheme to regularize the divergences appearing in the vacuum part of Eq.~(\ref{A3}). We here adhere to regularizing the integrals by introducing the 3-momentum cut-off $\Lambda$. The mass gap equation is obtained by minimizing $\Phi^{\rm NJL}_{\rm G}$ in Eq.~(\ref{A3}) with respect to the constituent mass $M$ at each space point~\cite{Wang:2018sur}, i.e.,  
	\begin{eqnarray}
		&& \frac{\del \Phi^{\rm NJL}_{\rm G}}{\del M}=0,\label{A4}\\
		\text{with the constraint, } &&\frac{\del^{2} \Phi^{\rm NJL}_{\rm G}}{\del M^{2}}> 0.\label{A5}
	\end{eqnarray}
	Using Eq.~(\ref{A4}), one obtains the following explicit form of the gap equation for the constituent mass $M$~\cite{Chen2021},
	\begin{eqnarray}
		M=m&+&\frac{G_{S}N_{c}N_{f}}{4\pi^{2}}\sum_{k=-\infty}^{\infty}\int_{0}^{\Lambda} dp_{\perp}^{2} \int_{-\sqrt{\Lambda^{2}-p_{\perp}^{2}}}^{\sqrt{\Lambda^{2}-p_{\perp}^{2}}} dp_{z}\left(J^{2}_{k}(p_{\perp}\rho)+J^{2}_{k+1}(p_{\perp}\rho)\right)\frac{M}{E}\nn\\
		&-&\frac{G_{S}N_{c}N_{f}}{4\pi^{2}}\sum_{k=-\infty}^{\infty}\int dp_{\perp}^{2} \int dp_{z}\left(J^{2}_{k}(p_{\perp}\rho)+J^{2}_{k+1}(p_{\perp}\rho)\right)\frac{M}{E}\left[\frac{1}{e^{\bt(E-(k+\frac{1}{2})\Om)}+1}+\frac{1}{e^{\bt(E+(k+\frac{1}{2})\Om)}+1}\right].\nonumber\\
		\label{A6}
	\end{eqnarray}
\end{widetext}
One has to solve Eq.~(\ref{A6}) along with the constraint~(\ref{A5}) to get the constituent quark mass as a function of angular velocity ($\Omega$), radius ($\rho$), temperature ($T$), 3-momentum cut-off ($\Lambda$), and the coupling strength ($G_{S}$). However, we can fix the 3-momentum cut-off $\Lambda$ and the coupling strength $G_{S}$ by using the pion mass and decay constant in vacuum~\cite{Buballa:2003qv,Klevansky:1992qe}, so that $M=M(\rho,\Omega,T)$. Using the constituent quark mass in Eqs.~\eqref{Shear_Rel},~\eqref{Isoshe}, and \eqref{new6} and appropriate degeneracy factors, we obtain the following expressions of shear viscosities and conductivities for the two-flavor rotating system,
\begin{eqnarray}
	&&\eta^{\rm NJL}_{0}=\eta^{\rm NJL}=\frac{8}{5T}  \int\frac{d^3p}{(2\pi)^3}\tau_{c}\frac{p^4}{E^2}f^{0}(1-f^{0})~,\label{A20}\\
	&&\eta^{\rm NJL}_{\perp,||,\times}=\frac{8}{5T}\int\frac{d^3p}{(2\pi)^3}\tau_{\eta\{\perp,||,\times\}}\frac{p^4}{E^2}f^{0}(1- f^{0}),
	\label{A21}\\
	&&\sigma^{\rm NJL}_{||} =\sigma^{\rm NJL}=\frac{20e^{2}}{9T}\int \frac{d^{3}p}{(2\pi)^3}\tau_{c} \frac{p^2}{E^{2}}f^{0}(1-f^{0}),\label{new10}\\
	&&\sigma^{\rm NJL}_{\perp,\times} =\frac{20e^{2}}{9T}\int \frac{d^{3}p}{(2\pi)^3}\tau_{\sigma\{\perp,\times\}} \frac{p^2}{E^{2}}f^{0}(1-f^{0}),\label{new11}
\end{eqnarray}
where we have multiplied an extra factor of two with the spin and color degeneracy factor $g_{r}$ to incorporate the anti-quarks, which contribute equally to $\mu=0$. In Eqs.~\eqref{A20} to \eqref{new11} the quasiparticle quark energies $E(\rho,\Omega,T)=\sqrt{p^{2}+M^{2}(\rho,\Omega,T)}$ are used in the distribution $f^{0}=1/[e^{-E(\rho,\Omega,T)/T}+1]$.

\section{Results and Discussion}
\label{Sec:Results_Discussion}

In this section, our primary objective is to investigate the influence of rotation, particularly the Coriolis force, on the shear viscosity and the electrical conductivity 
in two different model frameworks:
(i) the combined massless QGP (for $T>T_c$) and HRG (for $T<T_c$) framework, and
(ii) the NJL framework. 
For the first setting, we estimate the isotropic, parallel, perpendicular, and Hall components of shear viscosity for the hadronic phase using Eqs.~\eqref{Heta0} and \eqref{Heta1} and for the QGP phase using Eqs.~\eqref{isoQ} and \eqref{shQ} given in Sec.~\ref{HRGS}. Similarly, for electrical conductivity, the parallel (which is the same as isotropic component), perpendicular, and the Hall components in the hadronic phase is obtained using Eqs.~\eqref{new7} and \eqref{new8}, and in the QGP phase using Eqs.~\eqref{isosiQ} and \eqref{siQ} given in Sec.~\ref{HRGS}. The expressions of the transport coefficients within the NJL model are given in Eqs.~\eqref{A20}-\eqref{new11} of Sec.~\ref {NJLS}. 
In both settings, we use the same model of relaxation time $\tau_c$, i.e., a temperature-independent, constant $\tau_c$ for $T>T_c$, and a hard sphere-scattering calculated $\tau_c$ with a fixed radius for $T<T_c$. The advantages of this specific model for the relaxation time will be explained further in this section.

\begin{figure}[h]
	\centering
	\includegraphics[width=\columnwidth]{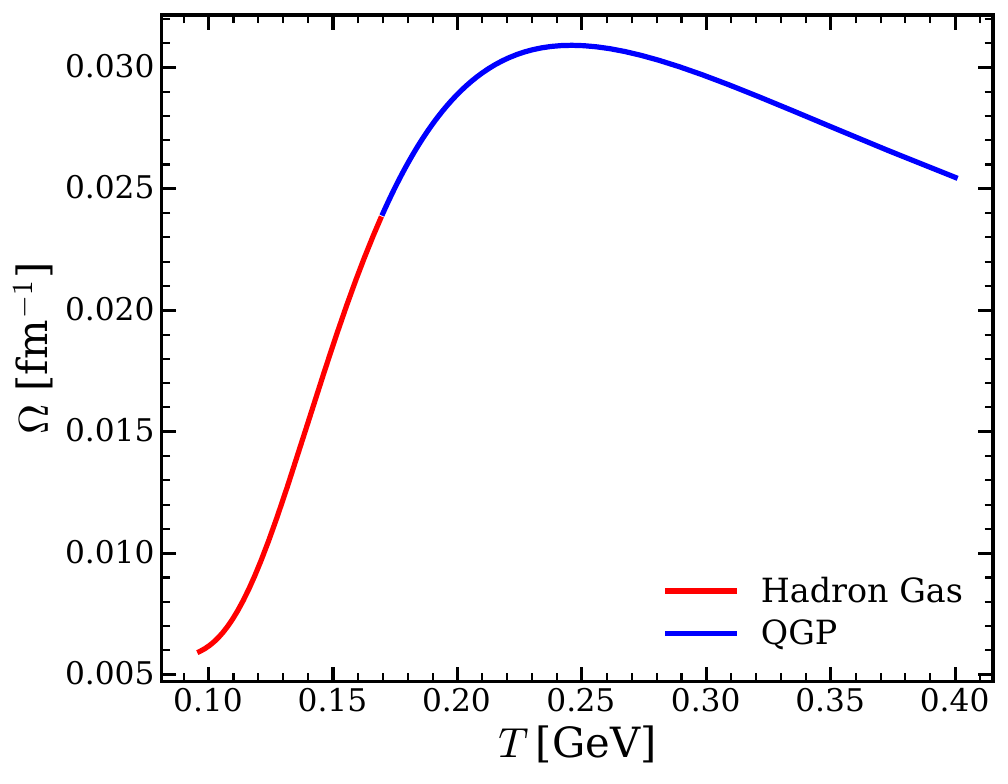}
	\caption{(Color online)  The variation of angular velocity with temperature.}
	\label{Figure_omega}
\end{figure}

The vorticity or angular velocity $\Omega$ in the medium created in HIC is not a constant, but rather space-time dependent. We can approximately map the temperature dependency of $\Omega$ using hydrodynamic evolution as follows. In HIC experiments, some fraction of initial OAM gets transferred to the medium in the form of local vortices, and the medium acquires a non-zero average angular velocity, $\Omega$. Since the medium expands with time, its moment of inertia ($I$) increases, and therefore, the average angular velocity $\Omega$ should decrease to keep the angular momentum $L\sim I\Omega$ conserved. Moreover, the expansion leads to cooling in time, which can be described by hydrodynamic theories. Using the cooling rate (temperature dependence on time) from a hydrodynamic model, we can map the temperature dependence of $\Omega$.
First, for the time-dependent angular velocity profile, we use the parametrization given by Jiang \textit{et al.} \cite{Jiang:2016woz} using a multiphase transport (AMPT) model simulations\footnote{In Refs.~\cite{Jiang:2016woz}, $y-$ axis is chosen perpendicular to the reaction plane whereas we have defined $z-$ axis to be perpendicular to the reaction plane (cf. Fig.~\ref{fig:2a}).}, 
\begin{align}
	\Omega(t, b, \sqrt{s_{NN}}) &=\frac{1}{2} \bigg[A(b, \sqrt{s_{NN}}) + B(b, \sqrt{s_{NN}}) \nn\\ 
	& \times (0.58t)^{0.35} e^{-0.58t}\bigg],\label{vort}  
\end{align}
where $\sqrt{s_{NN}}$ is the center of mass energy of colliding beams, $b$ is the impact parameter. The coefficients \( A \) and \( B \) are given by  
\begin{align}
	A &= \left[ e^{-0.016 b \sqrt{s_{NN}}} + 1 \right] \times \tanh(0.28 \, b)\nn\\
	& \times \left[ 0.001775 \tanh(3 - 0.015 \sqrt{s_{NN}}) + 0.0128 \right], \\
	B &= \left[ e^{-0.016 b \sqrt{s_{NN}}} + 1 \right] \times \left[ 0.02388 \, b + 0.01203 \right]\nn\\
	& \times \left[ 1.751 - \tanh(0.01 \sqrt{s_{NN}}) \right].
\end{align}
where \( \sqrt{s_{NN}} \) is measured in GeV, \( b \) in fm, the time \( t \) in fm/c, and \( \Omega \) in fm\(^{-1}\). Second, to obtain a temperature-dependent map for $\Omega$, we consider the simple hydrodynamic model introduced by Bjorken. In this model, the cooling rate is obtained using $(1+1)$-dimensional longitudinal expansion as~\cite{PhysRevD.27.140}
\begin{equation}
	T(t) = T_0 \left( \frac{t}{t_0} \right)^{-1/3}~.\label{bjorken}
\end{equation}
In this analysis, we assume beam energy $\sqrt{S_{NN}}= 200$~GeV with an impact parameter \(b=\) 5 fm. The initial conditions for hydrodynamic evolution are taken as --  initial temperature \(T_0=0.4\) GeV and initial time \(t_0=\) 0.14 fm/c. The corresponding $\Omega(T)$ profile obtained using Eqs.~\eqref{vort} and \eqref{bjorken} is displayed in Fig.~\eqref{Figure_omega}. 

In the following, we discuss the estimation of transport coefficients using both a constant $\Omega$ and the temperature-dependent $\Omega(T)$ obtained above.
We divide the results into two subsections describing the transport coefficients in the QGP--HRG framework in \ref{resultsHRG} and then the NJL framework in \ref{resultsNJL}. 

\subsection{Results for the shear viscosity in the QGP--HRG framework}\label{resultsHRG}
	
\begin{figure*}[!t]
	\centering
	\begin{subfigure}[b]{0.32\textwidth}
		\centering
		\includegraphics[width=\linewidth]{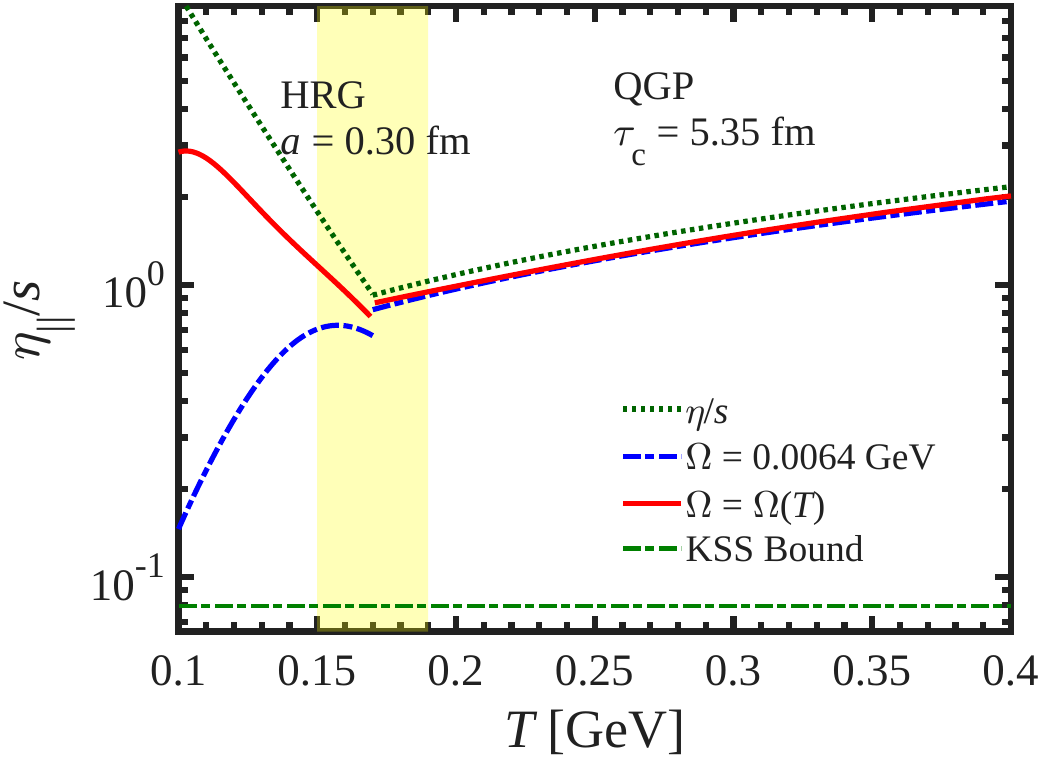}
		\caption{Parallel component.}
		\label{Figure_2a}
	\end{subfigure}
	\hfill
	\begin{subfigure}[b]{0.32\textwidth}
		\centering
		\includegraphics[width=\linewidth]{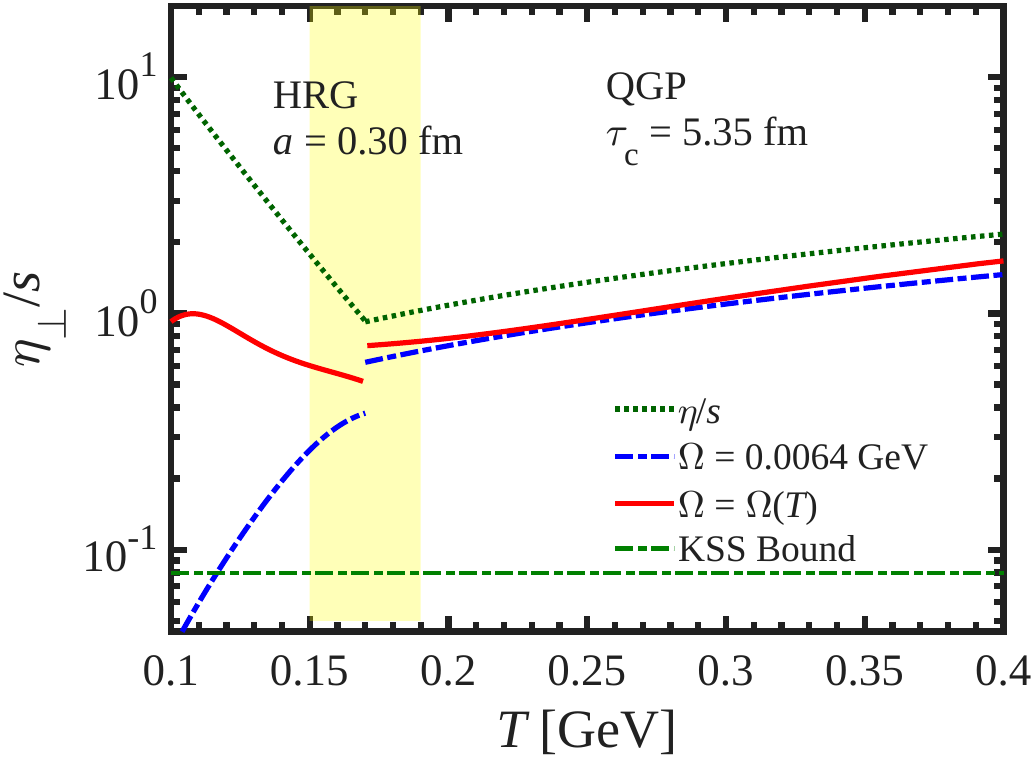}
		\caption{Perpendicular component.}
		\label{Figure_2b}
	\end{subfigure}
    \hfill
    \begin{subfigure}[b]{0.32\textwidth}
		\centering
		\includegraphics[width=\linewidth]{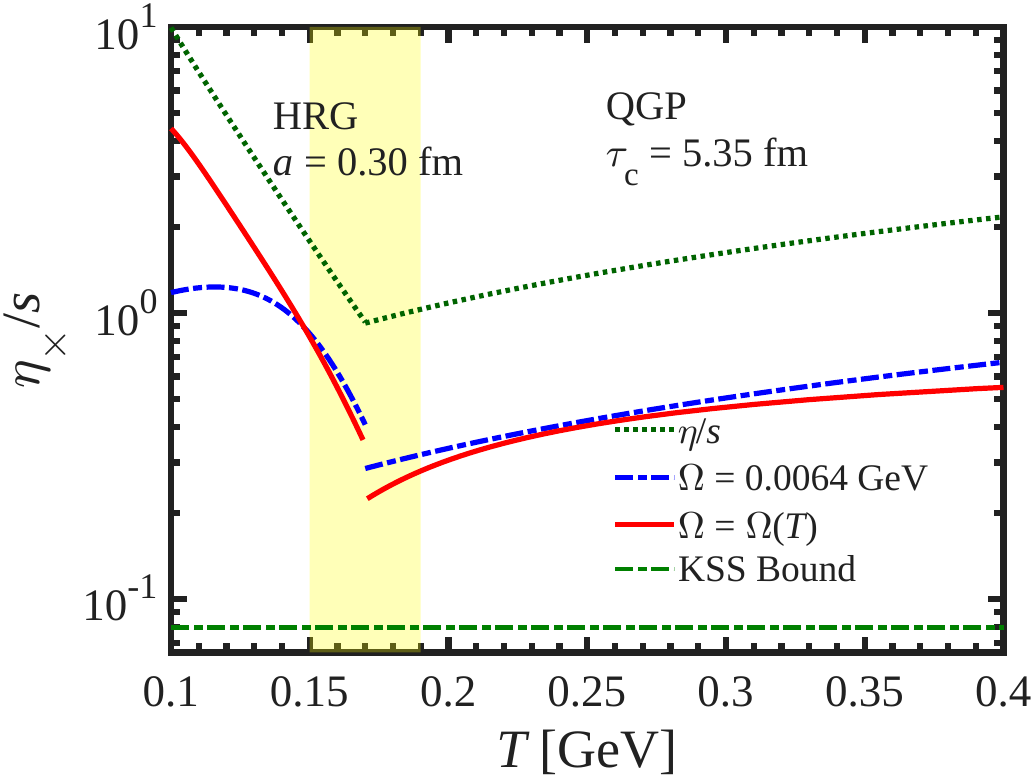}
		\caption{The Hall component.}
		\label{Figure_2c}
	\end{subfigure}
	
	\caption{Ratio of anisotropic shear viscosity components to entropy density as a function of temperature for a constant angular velocity
    $\Omega = 0.0064$~GeV (blue dash-dot curves) and a temperature-dependent angular velocity
    $\Omega = \Omega(T)$ (red solid curve), compared with the isotropic shear viscosity in the absence of rotation (green dotted curves). The critical temperature is taken to be 0.17~GeV.}
	\label{Figure_2}
\end{figure*}

For our investigation of shear viscosity within rotating hadronic matter, we employ the HRG model, incorporating a comprehensive spectrum of hadrons and their resonances with masses up to 2.6 GeV~\cite{ParticleDataGroup:2008zun}, while we consider the rotating QGP phase to be a massless gas of quarks and gluons. We note that the electrical conductivity within this framework has already been studied by some of us in Ref.~\cite{Padhan:2025nps}. Therefore, we only summarize those results qualitatively, rather than reproducing the corresponding figures, and discuss them in connection with the shear viscosity wherever relevant. Before discussing the trend of shear viscosity in the presence of angular velocity, let us first analyze the behavior of the shear viscosity in the absence of rotation. This serves as a reference for the parametrization of the relaxation time. The theoretical calculations~\cite{Gorenstein:2007mw, Kadam:2014cua,Itakura:2007mx,Fernandez-Fraile:2009eug,Plumari:2012ep,Lang:2012tt,Marty:2013ita,Noronha-Hostler:2008kkf} of shear viscosity suggest a valley-shaped $\eta/s$ as a function of temperature with a dip around critical temperature $T_{c}$ (chosen here as 170 MeV). On the other hand, the shape of the $\eta/s$ in the quasi-particle kinetic theory approach employed here mainly depends on two competing factors--thermodynamic phase space (P.S.) and relaxation time. For a system in the absence of any external field and for a momentum-independent relaxation time, we can write $\eta^{\rm  HRG}/s^{\rm HRG}=\frac{1}{s^{\rm HRG}}\sum_{B,M}({\rm P.S.})_{\eta}^{B,M}\tau_{c}^{B,M}$, where $({\rm P.S.})_{\eta}^{B,M}=\frac{g_{B,M}}{15T}\int \frac{d^{3}p}{(2\pi)^{3}}\frac{p^{4}}{E_{B,M}^{2}}f^{0}_{B,M}(1\pm f^{0}_{B,M})$. In the QGP phase, the same relation holds, which can be easily ascertained from Eqs.\eqref{isoQ} and \eqref{Qent}. The P.S. part of $\eta/s$ increases slowly with temperature, and the relaxation time dominates the temperature dependence of $\eta/s$. In the present paper, the valley-shaped behavior of $\eta/s$ is mimicked by modeling the relaxation time. A constant relaxation time in the QGP phase ($T>T_{c}$) and a hard sphere-scattering type relaxation time in the hadronic phase ($T<T_{c}$) are chosen to serve our purpose. This way of parameterizing $\tau_{c}$ in two different regions (quark matter and hadronic matter) is motivated by its simplicity and ability to cover the model calculations. Several papers~\cite{Dwibedi:2024mff,Dwibedi_2025,Padhan:2025nps,Dwibedi:2025fnz} have employed a similar way of modeling $\tau_{c}$ in the context of HICs. The theoretical calculation of electrical conductivity also suggests a valley-like shape, and the same parametrization procedure can be applied to it. Readers can refer to Fig.(1) of Reference~\cite{Padhan:2025nps} to see the calibration of $\tau_{c}$ for the electrical conductivity. 
In the current work, the calibration of $\tau_{c}$ for the estimation of $\eta/s$ has been performed in Figure~\eqref{Figure_1} of appendix~\ref{appeta_cal}. Using the calibrated relaxation time and scattering length (or hard core radius) given in Fig.~(\ref{Figure_1}), we will describe the effect of Coriolis force and anisotropic components of shear viscosity ($\eta_{||,\perp,\times}$) in  Fig.~(\ref{Figure_2}). Before this, let us note that, in the presence of angular velocity, it is again the effective relaxation times that dictate the behavior of the anisotropic components of shear viscosity. This can be easily seen by expressing the anisotropic viscosity components given in Eq.~\eqref{Heta1} for the HRG phase in a compact form,    
\begin{eqnarray}
	&& \eta^{\rm HRG}_{\perp, ||, \times}(T)=\sum_{B,M} \tau_{\eta \{\perp,||,\times\}}^{B,M} ({\rm P.S})_\eta^{B,M}~,\nonumber
\end{eqnarray}
where the phase space part only depends on temperature; on the other hand, the relaxation part has a different structure for different components of the viscosity. The effective relaxation times that modulate the thermal relaxation time $\tau_{c}(T)$ are given by Eq.~\eqref{relaxH}. Similarly, for components of the shear viscosities in the QGP phase, the relaxation part and phase space part are easily seen from Eq.~\eqref{shQ}.

Fig.~(\ref{Figure_2}) depicts the temperature dependence of the anisotropic components of the shear viscosity-to-entropy density ratio (\(\eta_{||,\perp,\times}/s\)) under rotation in both the hadronic and QGP phases. To depict the variation of viscosity in the hadronic phase, we choose a scattering length of \(a = 0.30\) fm, which falls in the band of scattering length obtained in Fig.~(\ref{Figure_1}). In the QGP phase, relaxation time \(\tau_c=5.35\) fm is chosen. The impact of rotation is analyzed for two cases: a constant angular velocity (\(\Omega = 0.0064\) GeV) and a temperature-dependent angular velocity \(\Omega = \Omega(T)\). Now, let us analyze the variation of anisotropic shear viscosity-to-entropy density ratio (\(\eta_{||,\perp,\times}/s\)) as a function of temperature as displayed in Fig.~(\ref{Figure_2}). For comparison, each plot also includes the isotropic shear viscosity component, $\eta\equiv \eta_{0}$, shown by the green dotted curve. The constant angular velocity $\Omega=0.0064$ GeV$=0.0325$ fm$^{-1}$ (or, $ \tau_{\Omega}=15.39$ fm) used in this analysis represents the maximum possible angular velocity provided by Eq.~(\ref{vort}) for $\sqrt{S_{NN}}=200$ GeV with $b=5$ fm. 

From Fig.~(\ref{Figure_2a}), in the QGP phase ($T>0.17$ GeV), the parallel component $\eta_{||}/s$ with fixed angular velocity $\Omega=0.0064$ GeV (blue dash-dot curve) lies slightly below the corresponding result with temperature-dependent angular velocity $\Omega(T)$ (red solid curve). Both anisotropic results are marginally lower than the isotropic $\eta/s$ (green dotted curve). In the hadronic phase, the ordering of the viscosity magnitudes are same as of the QGP phase, i.e., $\eta_{||}(\Omega=0.0064)/s$ (blue dash-dot curve) $<$ $\eta_{||}(\Omega(T))/s$ (red solid curve) $<$ $\eta/s$ (green dotted curve) with significant difference in their magnitudes. These results can be mathematically understood by comparing the rotational time scale $\tau_{\Omega}(T)$ with the relaxation time scale $\tau_{c}(T)$, which are present in the expressions of $\eta_{||}$. 
The effective relaxation times $\tau_{\eta ||}(\Omega=0.0064)<\tau_{\eta ||}(\Omega(T))$ make the blue dash-dot curve lie below the red solid one. Fig.~(\ref{Figure_2b}) describes the same physics for the perpendicular component of viscosity. A comparison with $\eta_{||}/s$ suggest that the magnitude $\eta_{\perp}/s$ is lesser than $\eta_{||}/s$. This trend can also be understood from their respective expression~\eqref{Heta1}, which implies $\eta_{||}(T)/s>\eta_{\perp}(T)/s$ as $\tau_{\eta||}>\tau_{\eta \perp}$. Fig.~(\ref{Figure_2c}) depicts the variation of Hall viscosity $\eta_{\times}/s$ with respect to temperature. One can observe a non-zero Hall component of viscosity, even for net baryon-free matter, both in the QGP and HRG phases, as a result of rotation. 

In the temperature-dependent variation of $\eta_{||,\perp,\times}/s$, we observed QGP-phase viscosity components for constant $\Omega=0.0325$ fm$^{-1}$ (corresponding to blue dash-dot curve) and for the temperature-dependent $\Omega(T)$ (red solid curve) nearly coincide, whereas a notable difference is seen in the hadronic phase. This behavior can be understood by falling back to the expressions~\eqref{vort} and~\eqref{bjorken}. These equations indicate a smaller band of $\tau_{\Omega}\sim 15.39-20$ fm for the QGP temperature range $T=0.4-0.17$ GeV but a larger band of $\tau_{\Omega}\sim 20-84$ fm for the HRG temperature domain $T=0.07-0.1$ GeV. To summarize, for all three plots shown in Fig.~(\ref{Figure_2}), the key factor determining the shape and magnitude of the anisotropic viscosities is the ratio between the average thermal relaxation time $\tau_{c}$ and rotational time scale $\tau_{\Omega}$.

\subsection{Results for the shear viscosity and electrical conductivity in the NJL framework}\label{resultsNJL}
In this section, we describe the behavior of shear viscosity and electrical conductivity obtained in the NJL model. Before discussing these transport coefficients, we first describe the main input, which modifies these transport parameters, i.e., the constituent quark mass $M(\rho,\Omega,T)$.    
\begin{figure}[H]
	\centering
	\includegraphics[width=\columnwidth]{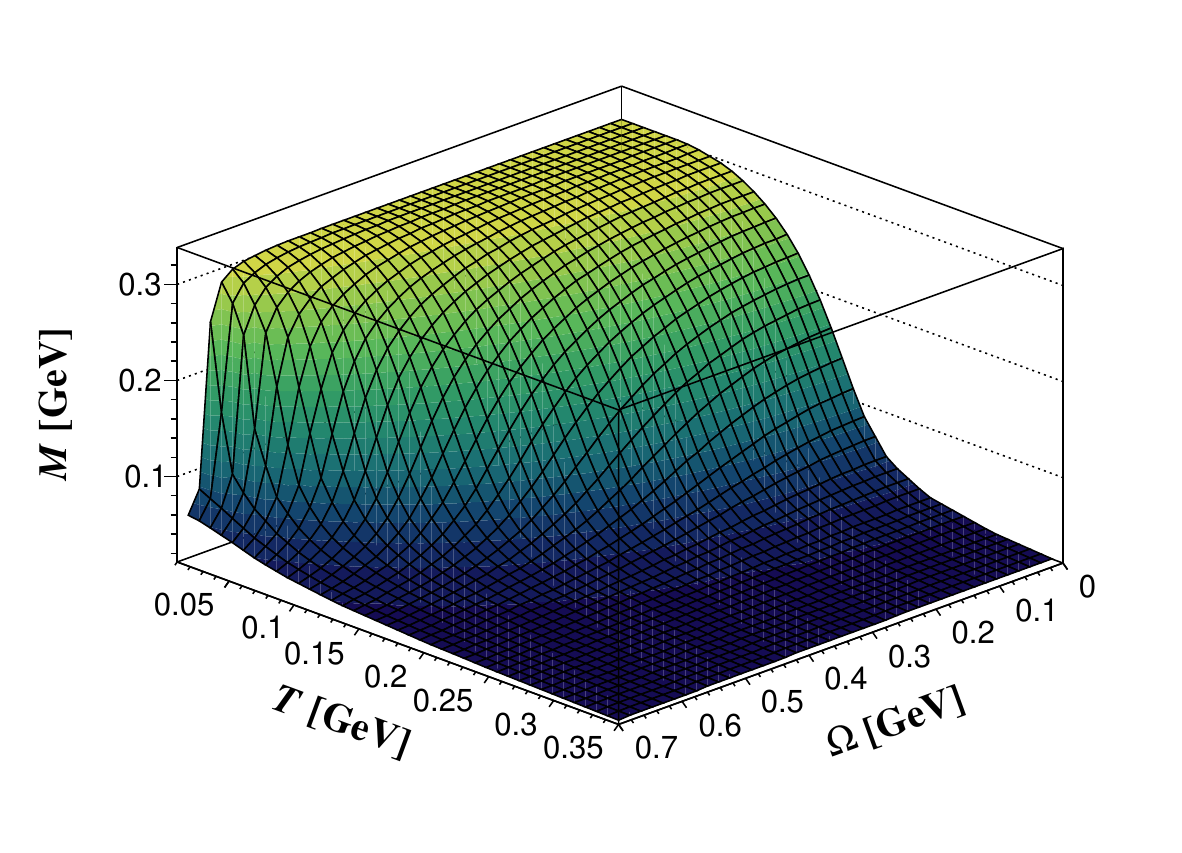}
	\caption{Constituent quark mass ($M$) as a function of angular velocity ($\Omega$) and temperature ($T$).}
	\label{fig:Mass}
\end{figure}
For the numerical estimation of constituent quark mass and thermodynamic equation of state (EoS) from the NJL model, we used the following parameter set. To fit pion mass and decay constant in the vacuum, we consider current quark mass $m_u = m_d = 5.5$~MeV, cut off to regularize the vacuum term is $\Lambda = 651$~MeV, and the scalar coupling $G_s = 5.04 \times 10^{-6}$~MeV$^{-2}$~\cite{Klevansky:1992qe}. Moreover, all the results are at position $\rho= 0.1$~GeV$^{-1}$~\cite{Jiang:2016wvv}, and for vanishing baryon chemical potential, $\mu_B = 0$. Angular velocity $\Omega$ is a free parameter bounded by the causality relation $\rho \Omega < 1$. 

In Fig.~(\ref{fig:Mass}), we depict constituent quark mass ($M$) as a function of angular velocity ($\Omega$) and temperature ($T$). As the temperature $T$ increases, the value of the constituent quark mass reduces from $M=313$~MeV to $M=m_{u}=m_{d}=5.5$ MeV. This is the standard representation of a chiral phase transition from the chiral symmetry-breaking phase to the restored phase. The temperature where the change of mass ($\frac{dM}{dT}$) becomes maximum is commonly known as the chiral phase transition temperature $T_{c}$. Interestingly, we notice that $M$ also reduces in the rotating frame with increasing $\Omega$. As a result, rotation can reduce the chiral transition temperature ($T_c$). However, in HIC, the vorticity or $\Omega$ is substantially low ($\approx 0.01$~GeV~\cite{XuGuangHuang:2020dtn,PhysRevC.95.054915,XuGuangHuang2016,Jiang2016}) to make any significant impact in $T_c$.  Even if we increase $\rho$, as shown in Ref.~\cite{Jiang:2016wvv}, at such low $\Omega$, effect in $M$ or $T_c$ is negligible. Though phenomenological values of $\Omega$ are not high enough to have an impact on $T_{c}$, different thermodynamical and transport quantities can be modified due to the temperature dependence of $M$. We will see these quantities in the next figures.

\begin{figure}[!t]
	\centering
	\includegraphics[width=\columnwidth]{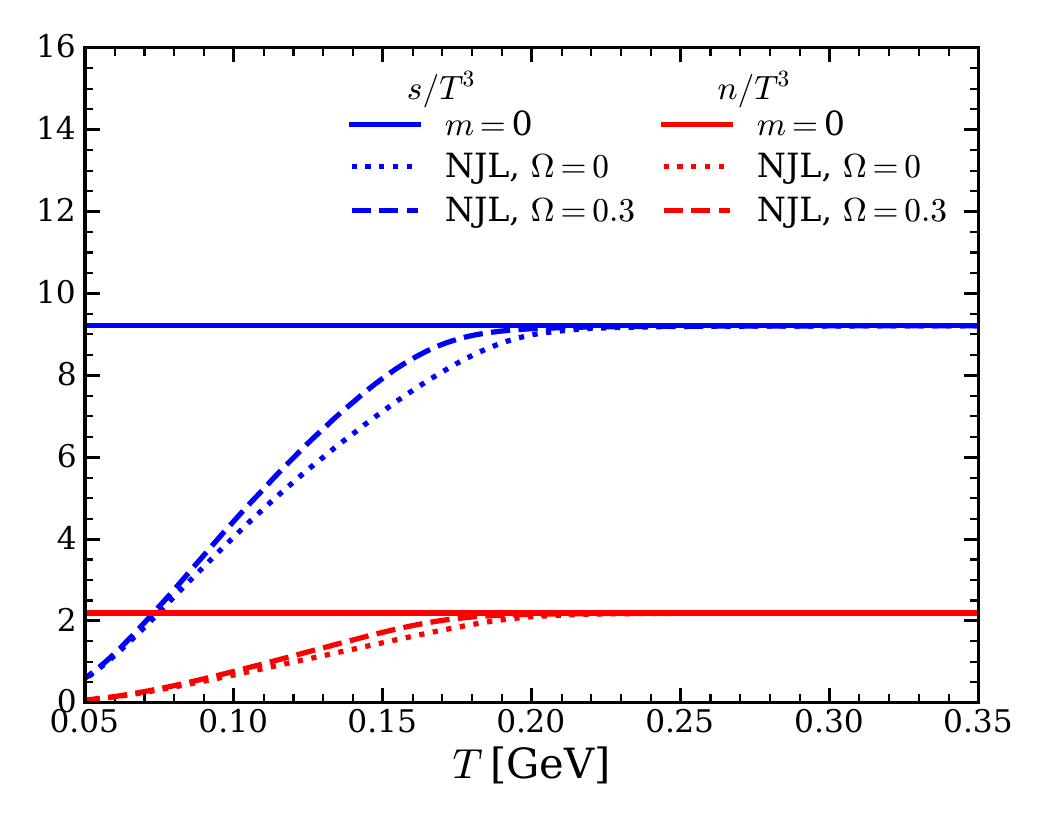}
	\caption{Variation of thermodynamic variables: entropy $s$ (blue) and total number density $n$ (red) as a function of temperature $T$ at different values of $\Omega$ in GeV.}
	\label{fig:Thermo}
\end{figure}
In Fig.~(\ref{fig:Thermo}), we plot the entropy density ($s$) and total quark number density ($n$) scaled by $T$ as a function of temperature at two different values of $\Omega = 0, 0.3$ GeV. The dotted and dashed lines represent NJL model estimations obtained from the grand potential given in Eq.~(\ref{A3}) and with the help of the Euler thermodynamic relation. Solid lines represent results for a massless, non-interacting gas of quarks. We can see that at a higher angular velocity, the thermodynamic quantities are enhanced (dashed lines lie above the dotted ones). This is a result of a decrease in quark condensate with $\Omega$. Besides thermodynamic EoS, the behavior of $s$ is important to analyze the fluidity measure, which is a ratio of shear viscosity to entropy density ($\eta/s$). On the other hand, the total quark number density $n$ is an important quantity for the determination of the average relaxation time $\tau_{c}$ of the particles.
\begin{figure*}[!t]
	\centering
	\begin{subfigure}[b]{0.45\textwidth}
		\centering
		\includegraphics[width=\linewidth]{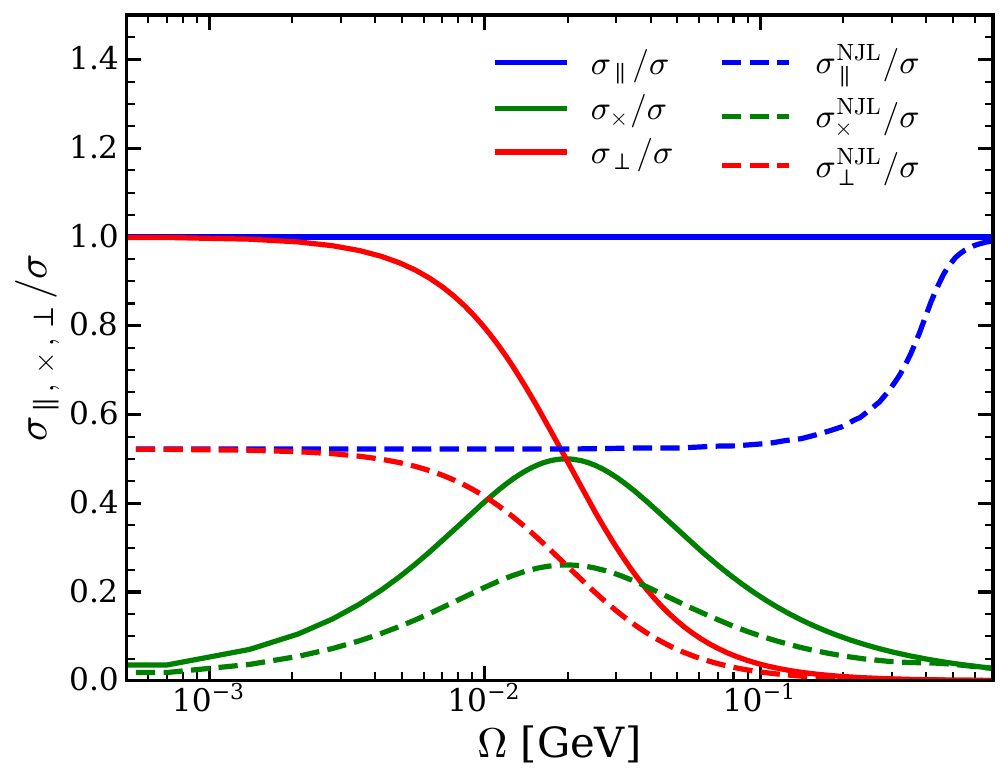}
		\caption{Electrical conductivity components against $\Omega$ at $T=150$ MeV and $\tau_{c}=5$ fm.}
		\label{nsigmaOm}
	\end{subfigure}
	\hfill
	\begin{subfigure}[b]{0.45\textwidth}
		\centering
		\includegraphics[width=\linewidth]{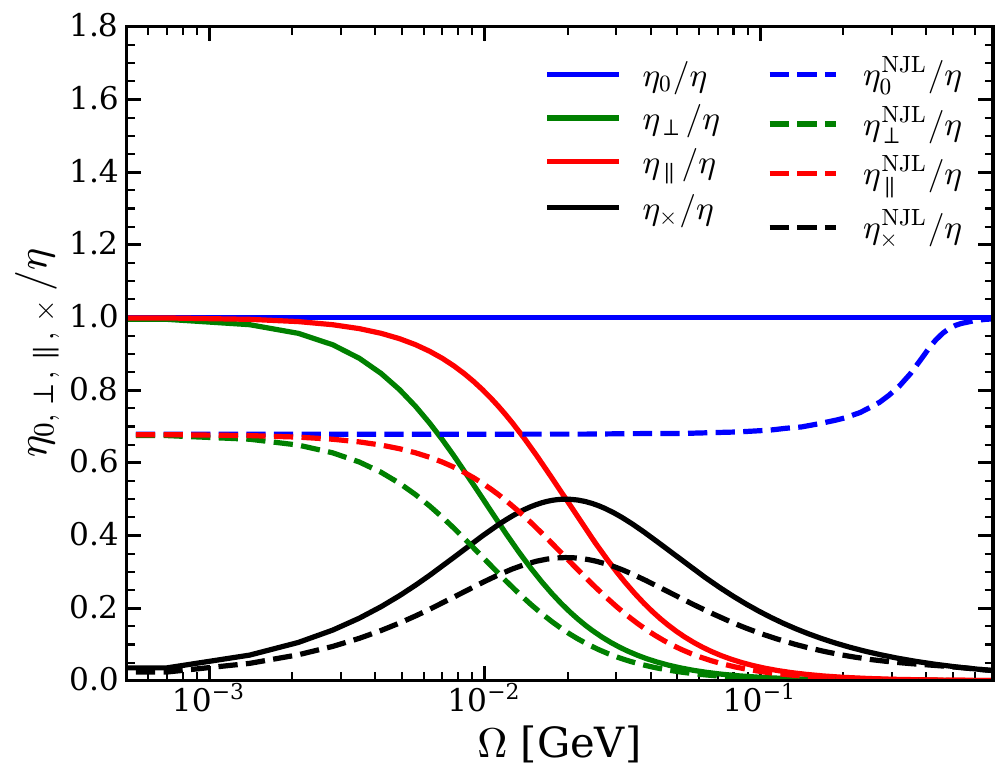}
		\caption{Shear viscosity components against $\Omega$ at $T=150$ MeV and $\tau_{c}=5$ fm.}
		\label{nsigmaT}
	\end{subfigure}
	
	\caption{Anisotropic transport coefficient components normalized to the corresponding isotropic coefficient, plotted as a function of angular velocity $\Omega$ at $T=150$ MeV. Here, $\sigma_{||}\equiv\sigma$ and $\eta_0\equiv\eta$ denote the isotropic conductivity and viscosity, respectively, in the non-rotating frame.}
	\label{fig:Nconductivities}
\end{figure*}
Fig.~(\ref{nsigmaOm}) shows the components of electrical conductivity $\sigma_{||,\perp,\times}$ in the rotating frame, normalized by the isotropic conductivity in the absence of $\Omega$ (here $\sigma= \sigma_{||}$, obtained for massless QGP). We plot their variation with angular velocity for a fixed temperature of $T = 150$~MeV. The parallel ($\parallel$), perpendicular ($\perp$), and Hall ($\times$) components are represented by the blue, red, and green lines, respectively. The dashed and solid lines are NJL and massless non-interacting estimations, respectively. 
 Here, we consider a constant relaxation time $\tau_c=5$ fm. By decreasing $\Omega$, the Hall-type component vanishes, and the other two components $\perp$ and $\parallel$ merge as a single component, which means the anisotropy of the conductivity tensor disappears (as expected). At low $\Omega$, massless results differ from the NJL estimation, which carries a non-zero constituent mass due to non-zero quark condensate. However, at very high $\Omega$, massless results recover due to chiral restoration. The Hall-type component in the rotating frame has a significant contribution at zero chemical potential, unlike in the presence of a magnetic field. The Lorentz force induced by a magnetic field generates anisotropic transport, with positively and negatively charged particles contributing oppositely to the Hall component. In contrast, the Coriolis force in a rotating frame does not distinguish particles based on their charge. As a result, Hall-type components can play a significant role in transport phenomena in rotating systems. The peak in the Hall component appears at a $\Omega$ value where rotational relaxation time ($\tau_\Omega$) equals thermal relaxation time, which here is $\tau_c =5$ fm, beacause of the factor $\frac{\tau_c/\tau_\Omega}{1 + (\tau_c/\tau_\Omega)^2}$ in the expression of $\sigma_{\times}$.   
Fig.~(\ref{nsigmaT}) represents the same as Fig.~(\ref{nsigmaOm}) but for shear viscosity components. There are few differences in the expressions of $\sigma_{||,\perp,\times}$ and $\eta_{||,\perp,\times}$, which can be understood from the earlier formalism section, e.g., $\sigma_{||}=\sigma$ but $\eta_{||}\neq \eta$, rather $\eta_{0}=\eta$. That is why we have drawn viscosity components by normalizing them with $\eta_{0}$ or $\eta$. 

\begin{figure*}[!t]
	\centering
	\begin{subfigure}[b]{0.45\textwidth}
		\centering
		\includegraphics[width=\linewidth]{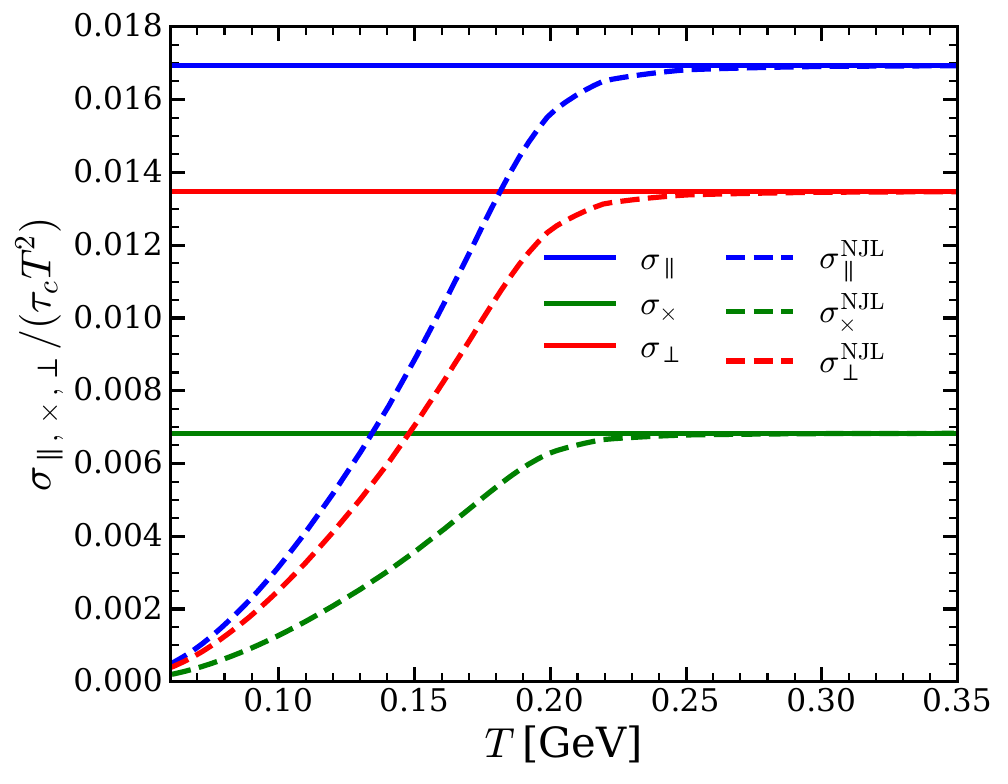}
		\caption{Electrical conductivity components vs $T$ at $\Omega=0.01$~GeV and $\tau_c=5$~fm.}
		\label{netaOm}
	\end{subfigure}
	\hfill
	\begin{subfigure}[b]{0.435\textwidth}
		\centering
		\includegraphics[width=\linewidth]{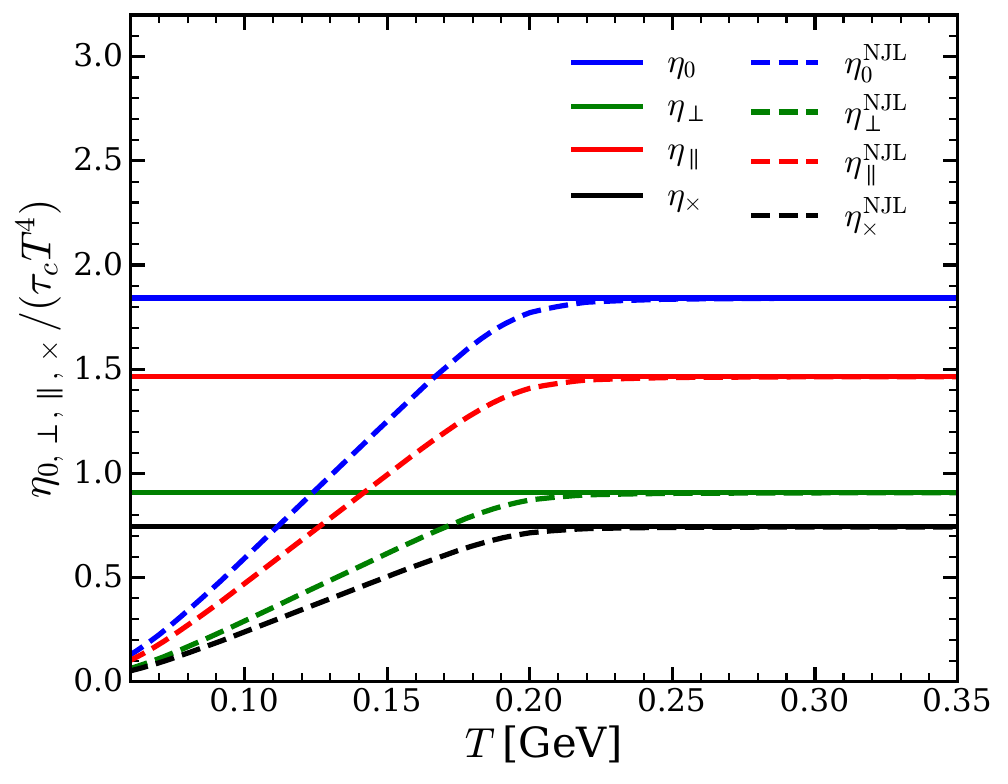}
		\caption{Shear viscosity components vs $T$ at $\Omega=0.01$~GeV and $\tau_c=5$~fm.}
		\label{netaT}
	\end{subfigure}
	
	\caption{Components of the transport coefficients, scaled by the relaxation time $\tau_c$ and temperature $T$, plotted as functions of $T$ at $\Omega=0.01$~GeV.}
	\label{fig:Nviscosities}
\end{figure*}
Fig.~(\ref{netaOm}) depicts the variation of conductivity components with $T$ at $\Omega = 0.01$~GeV. Conductivity is made dimensionless by dividing $\tau_c T^2$. To highlight the effect of rotating thermodynamics, we set a fixed value of the relaxation time $\tau_c =5$~fm. The conductivity increases with $T$ as the constituent quark mass decreases. At high $T$, chirality restores, and the NJL estimations converge to the massless case. A similar trend is observed for the shear viscosity components, as shown in Fig.~(\ref{netaT}). We consider $\Omega=0.01$ GeV, which is within the phenomenological values. Here, we notice a substantial difference between parallel and perpendicular components (reflecting anisotropy) and a non-zero Hall component. 

In the above discussion, we observe that at high angular velocity ($\Omega\approx 10^{-1}$ GeV), the thermodynamics as well as transport variables can be affected. In particular, the shear viscosities have been observed to be enhanced at high $\Omega$. It is a result of partial restoration of chiral symmetry at high angular velocity, which is marked by the decrease in the constituent quark mass as a function of $\Omega$. However, the angular velocity range in the HIC experiments is expected to be very large, so that the effect can be observed (In fact, it is less by one order of magnitude, see Fig.~\eqref{Figure_omega}). This led us to focus on a realistic estimation of the transport coefficients of the rotating quark matter produced in HIC within the NJL model. To serve this purpose, we tune our relaxation time to cover the earlier estimations of the isotropic shear viscosity and conductivity in a manner similar to the previous section. The details of this calibration are provided in appendix~\ref{appeta_cal1}.
\begin{figure*}
	\centering
	\begin{subfigure}{.45\textwidth}
		\centering
		\includegraphics[width=\linewidth]{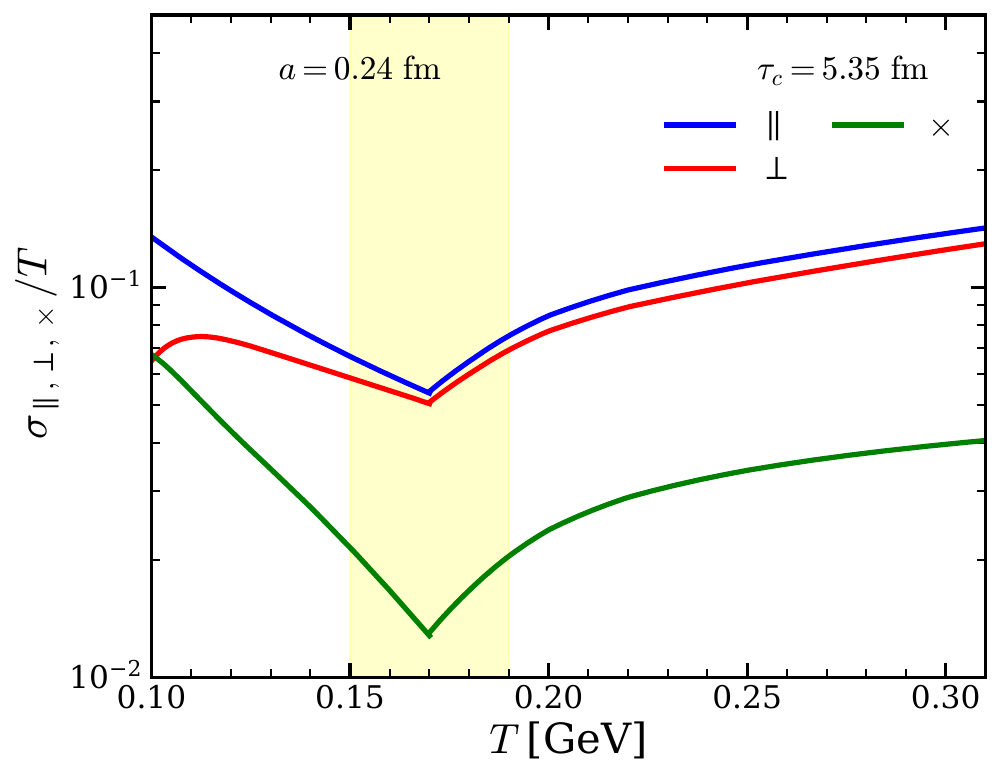}
		\caption{(Color online) The variation of anisotropic components of conductivity ($\sigma_{\perp,\times}/T$) against
			temperature by taking a temperature-dependent $\Omega(T)$ is compared with isotropic conductivity $\sigma_{||}/T$ in the absence of rotation.}
		\label{sigmaT}
	\end{subfigure}
	\hfill
	\begin{subfigure}{.48\textwidth}
		\centering
		\includegraphics[width=\linewidth]{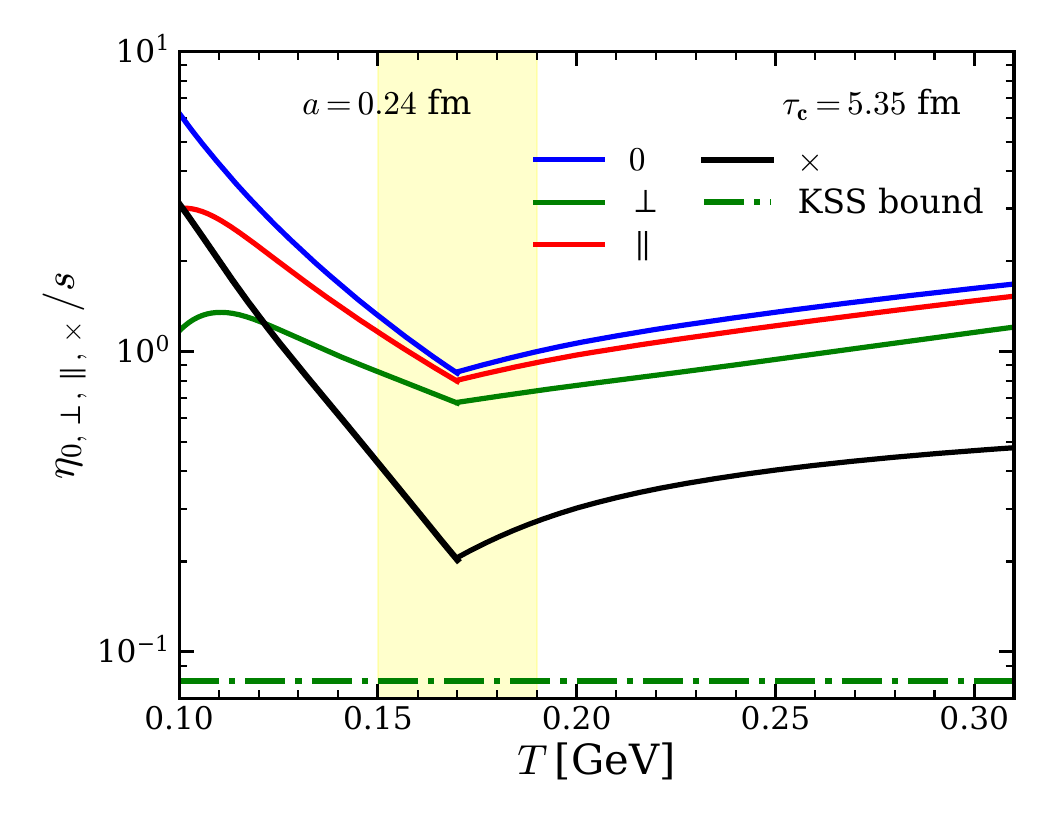}
		\caption{(Color online) The variation of anisotropic shear viscosity to entropy density ratio ($\eta_{\perp,||,\times}/T$) against
			temperature by taking a temperature-dependent $\Omega(T)$ is compared with isotropic component $\eta_{0}/s$ in the  absence of rotation.}
		\label{etaT}
	\end{subfigure}
	\caption{Estimation of $\sigma/T$ and $\eta/s$ using phenomenological $\Omega(T)$ for matter created in HIC.}
	\label{fig:sigma_eta}
\end{figure*}
In Fig.~(\ref{sigmaT}) and Fig.~(\ref{etaT}), we respectively show the variation of the scaled conductivities $\sigma_{\perp,\times}/T$ and shear viscosity to entropy density ratio $\eta_{\perp,||,\times}/s$ against temperature. We choose an intermediate value of $\tau_{c}=5.35$ fm for the quark temperature domain and scattering length $a=0.24$ fm for the hadron temperature domain, which falls in the band obtained in Figs.~(\ref{sigmatau}) and \eqref{etatau}. Readers should note that our choice of $\tau_{c}$ is the same as that of previous Sec.~\ref{resultsHRG}; however, we choose a different scattering length so that the NJL results are continuous across the critical temperature. The temperature-dependent $\Omega(T)$ is used in the expressions~(\ref{A20}) to~(\ref{new11}) for the computation of the anisotropic conductivity and viscosity. The isotropic component of conductivity ($\sigma_{||}$) and viscosity ($\eta_{0}$) have been shown for the purpose of comparison with other anisotropic components. In Fig.~(\ref{etaT}) we also display the conjectured KSS bound~\cite{Kovtun:2004de}, which is considered as a lower bound of the $\eta/s$. We observe in Fig.~(\ref{sigmaT}) and Fig.~(\ref{etaT}) that the perpendicular components of normalized viscosity and conductivity become smaller than their parallel component, which reflects the building of anisotropic transportation. On the other hand, the non-zero Hall components of transport coefficients are observed. Interestingly, all components follow the qualitative valley-shaped temperature profile due to the consideration of phenomenological $\Omega(T)$. In this regard, we have also noticed that these shapes can be spoiled by using a non-phenomenological constant $\Omega$ in the entire temperature range, which we do not show here (One can refer to Fig.~\eqref{Figure_2} of Sec.~\ref{resultsHRG} to see this for shear viscosity components in QGP--HRG framework). We can see a lesser anisotropy ($\sigma_{||}-\sigma_{\perp}$ or $\eta_{0}-\eta_{\perp,||}$) at higher temperatures ($T>170$ MeV) whose amount increases with decreasing temperature ($T<170$ MeV) and remain significant as the system approaches kinetic freeze-out ($T\approx 100$ MeV). The analogous behavior of the transport coefficients displayed in Fig.~\eqref{fig:sigma_eta} with that of Fig.~\eqref{Figure_2} is a result of the same model of the relaxation time chosen in both sections. Therefore, the reasoning of Sec.~\ref{resultsHRG} can be applied to justify the shape of the curves obtained in Fig.~\eqref{fig:sigma_eta}. For $T>T_{c}$, the thermodynamic phase space of different components $\eta/s$ is almost the same for the massless QGP and the NJL model. This is because the constituent mass of the quarks in the NJL model is negligible for temperature $T>T_{c}$ (see Fig.~\eqref{fig:Mass}). As a result, one should expect that the magnitude of different components of $\eta/s$ to be the same as that of corresponding NJL model estimates for $T>T_{c}$. This can be explicitly verified by comparing Figs.~\eqref{etaT} and \eqref{Figure_2}.

It is worth mentioning that our results at finite rotation are similar to those obtained at finite magnetic field \cite{Ghosh:2020wqx,Dey:2019axu}, and we can explore their phenomenological connections. In the RTA-based kinetic theory framework, the Coriolis force in the presence of rotation and the Lorentz force in the presence of magnetic fields induce anisotropy in the shear viscosity and electrical conductivity. In the present work, based on the QGP--HRG and NJL frameworks, we notice a significant anisotropy $\eta_{||}-\eta_\perp$ and $\sigma_{||}-\sigma_{\perp}$. The percentage reduction in the parallel and perpendicular components of viscosity $\left(\frac{\eta-\eta_{||,\perp}}{\eta}\times 100\%\right)$ and conductivity $\left(\frac{\sigma-\sigma_{\perp}}{\sigma}\times 100\%\right)$, for the temperature dependent $\Omega=\Omega(T)$ shown in Figs.~(\ref{Figure_2}) and \eqref{fig:sigma_eta} is as follows. In the domain of quark temperatures, the percentage reduction for massless QGP and NJL models is the same and is within the range $6-10\%$ for $\eta_{||}$ and $20-30\%$ for $\eta_{\perp}$. This is expected as constituent quark masses become negligible in this temperature domain as a result of chiral restoration. The reduction in the $\eta_{||}$ in QGP--HRG (NJL) framework is about $20-70\%$ ($10-50\%$) in the hadron temperature range, while the reduction in $\eta_{\perp}$ is about $45-90\%$ ($20-80\%$) in the hadron temperature range. The reduction in the $\sigma_{\perp}$ in the NJL framework is about $6-10\%$ in the quark temperature range and $6-50\%$ in the hadron temperature range (a comparable range is also obtained in the QGP--HRG framework in Ref.~\cite{Padhan:2025nps}). A similar reduction in $\eta_{||,\perp}/s$ and $\sigma_{\perp}/T$ is also noticed for the finite magnetic field picture. From the phenomenological point of view, as we move from head-on to peripheral collisions, we obtain a non-zero magnetic field and rotation. Our work suggests that the shear viscosity to entropy density ratio, as well as scaled conductivity, will have smaller values in peripheral collisions as compared to the corresponding estimates in head-on collisions. Lorentz force at finite magnetic field can lead to azimuthal anisotropy~\cite{Tuchin:2011jw,Mohapatra:2011ku}, and similar effects are expected due to Coriolis force at finite rotation. To investigate the phenomenology of the magneto-rotational effects, including azimuthal anisotropy, a systematic hydrodynamic evolution at finite magnetic field and rotation with the inclusion of all viscosity and conductivity components is required. Comparing experimental data from peripheral and head-on collisions may reveal signatures of a finite magnetic field and rotation. To isolate the effect of rotation, the Hall component can serve as a useful probe since it vanishes for $\mu = 0$ at finite magnetic fields. Moreover, photon or dilepton spectra are linked to the conductivity of the nuclear medium~\cite{Fernandez-Fraile:2009eug,Dwibedi:2025xho}, so one may also expect the anisotropies in the dilepton spectra because of the anisotropic conductivities. Our immediate future plan is to study in detail the above-mentioned phenomenological aspects of the anisotropy produced due to rotation.

\section{Summary and Conclusions}\label{Sec:Summary}
In the peripheral heavy-ion collision experiments, some fraction of initial angular momentum gets transferred to the formed quark-gluon plasma. By assuming a rotating plasma to account for this initial angular momentum, we set up a relativistic Boltzmann equation for the determination of its shear viscosities and electrical conductivities. The Coriolis force obtained from the non-trivial connection coefficients in the rotating frame enters the Boltzmann equation and makes the transport coefficients anisotropic. Using the Boltzmann equation in the rotating frame, we obtained five different components of shear viscosity and three different components of conductivity. The different components of the shear viscosity and conductivity obtained can be directly compared with the anisotropic shear viscosities and conductivities one finds in the presence of magnetic fields, where, in place of the angular velocity vector, the magnetic field vector breaks the isotropic nature of transport coefficients.
To get a realistic estimate of the shear viscosities and conductivities, we employ two different frameworks--the combined framework of massless quark-gluon plasma with the hadron resonance gas model and the Nambu--Jona-Lasinio model framework. In the case of the combined massless quark-gluon plasma and hadron resonance gas framework, we employed the hadron resonance gas model
below the critical temperature, whereas above the critical temperature, we took massless non-interacting partons as our degrees of freedom for the calculation of the viscosities and conductivities. In the Nambu--Jona-Lasinio framework, we obtain the constituent quark mass as a function of angular velocity and temperature. Subsequently, we explore the chiral symmetry restoration in the context of transport coefficients in the $\Omega$--$T$ plane with the help of constituent quark mass. Quark condensate in the NJL model with a fixed coupling constant value ($G_{S}$) reduces the pseudo-critical temperature with rotation. As a result, the constituent quark mass also decreases with rotation, and chiral restoration can be achieved at very high $\Omega$ even at low $T$. However, in a realistic range of $\Omega$ expected in heavy ion collisions, the modification of the constituent quark mass is solely due to the temperature of the medium. 

A compilation of previous microscopic estimates of the shear viscosity to entropy density ratio and the scaled electrical conductivity gives a well-known valley-like pattern against the temperature axis with a minimum around the transition temperature. By tuning the relaxation time in both frameworks, we have obtained the upper and lower curves that span these theoretical data. In both frameworks, we model the relaxation time of the system piece-wise, hard sphere scattering interactions with tunable scattering length in the hadronic temperature regime and a constant temperature-independent relaxation time in the quark temperature domain. Once tuned to match previous estimates of the isotropic viscosity-to-entropy density ratio and conductivity, the finite rotation extension gives a transformation from an isotropic to an anisotropic structure, characterized by three physical components for the viscosity: parallel, perpendicular, and Hall, and two physical components for the conductivity: perpendicular and Hall. The behavior of these anisotropic components is governed by two parts: an effective relaxation time that depends on both angular velocity and temperature, and a phase space factor that is purely temperature-dependent. The average vorticity or angular velocity of the nuclear matter mostly decreases as the system cools with time. To see the actual variation of shear viscosity and conductivity with temperature, we obtained the angular velocity as a function of temperature with the help of an approximate cooling law. In the same plot, we also showed the results with a constant angular velocity for comparison. We observe that the parallel and perpendicular components of the shear viscosity to entropy density ratio reduce due to rotation. The valley-like signature of the temperature dependence of shear viscosity is no longer valid if we consider a constant angular velocity throughout the evolution. However, with the realistic temperature-dependent angular velocity, the valley-like structure of the temperature dependence of the shear viscosity to entropy density ratio remains, though with a reduced magnitude. Unlike a magnetic field, rotation does not distinguish positively and negatively charged particles; as a result, the Hall-like transport component induced by rotation makes a significant contribution to transport phenomena. Moreover, a sufficient amount of anisotropy in the transport coefficients is noticed around kinetic freeze-out ($T\approx 100$ MeV), which can, in principle, affect the particle spectra. To the best of our knowledge, this work presents the first systematic characterization of the temperature dependence of shear viscosity and electrical conductivity components in a rotating medium, and provides a foundation for more detailed future studies. 

Finally, we would like to point out several directions for future work by relaxing some of the assumptions made in this study. In HICs, the produced medium can exhibit not only rotation but also significant radial flow. One possible approach to incorporate both effects is to perform an appropriate coordinate transformation that captures simultaneous rotation and radial expansion. In such a co-rotating and expanding frame, the resulting metric may differ from the form presented in Eq.~\eqref{S1}. This change in the metric can introduce additional pseudo-forces (cf. Eq.~\eqref{S2}), which may influence the transport phenomena. Moreover, the hard-sphere scattering approximation employed here assumes constant hadron radii for all the hadrons, which is a simplification. The hard-core radii of many baryons and mesons are known from electron–hadron scattering experiments~\cite{Cui:2022fyr,Fukushima:2020cmk}. From the HIC perspective, the hard-core radii are constrained from the lattice QCD data of the conserved charge susceptibilities~\cite{Vovchenko:2017xad,Karthein:2021cmb}. Incorporating species-dependent hard-core radii, rather than assuming the same value for all hadrons, could yield more accurate estimates of both the relaxation time and transport coefficients. Furthermore, there exists a more rigorous way of calculating the transport coefficients for the nuclear matter--the Green-Kubo technique~\cite{toda2012statistical}. This Green-Kubo method, combined with the hadronic transport approach such as SMASH (Simulating Many Accelerated Strongly interacting Hadrons)~\cite{Rose:2017bjz}, can provide transport coefficients of the system that do not even use the hard-core radii.

\section{Acknowledgement}
This work was supported in
part by the Ministry of Education, Government of India  (A.D., D.R.M., N.P.), Board of Research in Nuclear Sciences (BRNS) and Department of Atomic Energy (DAE), Government of India, with Grant Nos. 57/14/01/2024-BRNS/313 (S.G.), and the DAE-DST, Government of India funding under the mega-science project “Indian participation in the ALICE experiment at CERN” bearing Project No. SR/MF/PS-02/2021-IITI(E-37123) (K.G., R.S.).
K.G. acknowledges the financial support from the Prime Minister’s Research Fellowship (PMRF), Government of India.  
\appendix
\begin{widetext}
\section{Calculation of shear stress and conductivity tensors}\label{ape1}
In this appendix we explicitly solve Eq.~\eqref{S12} to get the unknown coefficients $C_{n}$ and $A_{n}$ needed for the calculation of shear stress and electric current provided in Eqs.~\eqref{S17} and \eqref{new5}. Let us begin with Eq.~\eqref{S12},
\begin{eqnarray}
	&&\frac{f^{0}(1+\xi f^{0})}{ET}\left(p^{i}p^{j}U^{ij} -q\tilde{E}^{i}p^{i}\right)+\frac{1}{\tau_{\Omega}}\ep^{ijk}p^j\om^{k} \frac{\partial \delta f}{\partial p^{i}} = -\frac{\delta f}{\tau_{c}}\nn\\
\implies &&	\frac{ p^i p^j}{ET} U^{ij} f^0(1+\xi f^0)-\frac{q\tilde{E}^{i}p^{i}}{ET}f^0(1+\xi f^0) +\frac{1}{\tau_\Om} \om^{ij} p^j \frac{\partial \delta f}{\partial p^{i}}= -\frac{\delta f}{\tau_{c}}~,\label{AA1}
\end{eqnarray}
where, $\tau_\Om\equiv1/2\Omega$ and $\ep^{ijk}\om^{k}\equiv \om^{ij}$. In terms of the unknown constants $C_{n}$ and $A_{n}$ we make the following guess for $\delta f$:
\begin{equation}
\delta f = \sum\limits_{n=0}^{6} C_{n}C_{n}^{kl}p^{k}p^{l}+\sum\limits_{n=0}^{2}A_{n}~A_{n}^{k}~p^{k}~.\label{AA2}
\end{equation}
The fluid field gradients $C_{n}^{ij}=C_{n}^{ijkl}U^{kl}$~\cite{Aung:2023pjf} and $A_{n}^{k}=A_{n}^{kl}\tilde{E}^{l}$~\cite{Padhan:2024edf} can be obtained from the contractions of the 4-rank tensor $C_{n}^{ijkl}$ and 2-rank tensor $A_{n}^{kl}$,
\begin{eqnarray}
	C^{ijkl}_0&=&(3\omega^i \omega^j-\delta^{ij})(\omega^k \omega^l-\frac{1}{3}\delta^{kl}),\nn\\
	C^{ijkl}_1&=&\delta^{il} \delta^{jk}+\delta^{jl} \delta^{ik}-\delta^{ij} \delta^{kl}+\delta^{ij} \omega^k \omega^l-\delta^{jl}\omega^i\omega^k\nn\\
	&-&\delta^{jk} \omega^i \omega^l+\delta^{kl} \omega^i \omega^j-\delta^{ik} \omega^j \omega^l-\delta^{il} \omega^j \omega^k+\omega^i \omega^j\omega^k \omega^l,\nn\\
	C^{ijkl}_2&=&\delta^{ik} \omega^j \omega^l+\delta^{il} \omega^j \omega^k+\delta^{jk} \omega^i \omega^l+\delta^{jl} \omega^i \omega^k-4\omega^i \omega^j \omega^k \omega^l,\nn\\
	C^{ijkl}_3&=&\delta^{il} \omega^{jk}+\delta^{jl} \omega^{ik}-\omega^{ik} \omega^j \omega^l-\omega^{jk} \omega^i \omega^l,\nn\\
	C^{ijkl}_4&=&\omega^{ik}\omega^j\omega^l+\omega^{il}\omega^j\omega^k+\omega^{jk}\omega^i \omega^l +\omega^{jl} \omega^i \omega^k,\nn\\ 
	C^{ijkl}_5&=&\delta^{ij}\delta^{kl},\nn\\
	C^{ijkl}_6&=&\delta^{ij}\omega^k\omega^l+\delta^{kl}\omega^{i}\omega^{j},\nn\\
	A^{kl}_{0}&=&\delta^{kl}\nn\\
	A^{kl}_{1}&=& \epsilon^{klj}\omega^{j}\nn\\
	A^{kl}_{2}&=&\omega^{k}\omega^{l}.\label{4-2tensor}
\end{eqnarray}
After contraction we can write,
\begin{eqnarray}
	C^{ij}_{0} &=& (3\omega^i \omega^j-\delta^{ij})(\omega^k \omega^l U^{kl}-\frac{1}{3} \vec {\nabla} \cdot \vec{u})~,\nn\\
	C^{ij}_{1} &=& 2U^{ij}+\delta^{ij}U^{kl}\omega^k \omega^l-2U^{ik}\omega^j \omega^k-2U^{jk} \omega^k \omega^i+(\omega^i \omega^j-\delta^{ij}) \vec{\nabla} \cdot \vec{u}+\omega^{i}\omega^{j}\omega^{k}\omega^{l} U^{kl}~,\nn\\
	C^{ij}_{2} &=& 2(U^{ik} \omega^j \omega^k+U^{jk} \omega^i \omega^k-2U^{kl}\omega^i \omega^j \omega^k \omega^l)~,\nn\\
	C^{ij}_{3} &=& U^{ik}\omega^{jk}+U^{jk}\omega^{ik}-U^{kl}\omega^{ik}\omega^j\omega^l-U^{kl}\omega^{jk}\omega^i \omega^l,\nn\\
	C^{ij}_{4} &=& 2(U^{kl} \omega^{ik} \omega^j \omega^l+U^{kl} \omega^{jk} \omega^i \omega^l)~,\nn\\
	C^{ij}_{5} &=& \delta^{ij}(\vec{\nabla}\cdot\vec{u})~,\nn\\
    C^{ij}_{6} &=& \delta^{ij}\omega^k \omega^l U^{kl}+\omega^i\omega^j (\vec{\nabla}\cdot\vec{u}),\nn\\
    A^{i}_{0}&=& \delta^{ij}\tilde{E}^{j}=\tilde{E}^{i},\nn\\
    A^{i}_{1}&=& \epsilon^{ijk}\omega^{k}\tilde{E}^{j}=\omega^{ij}\tilde{E}^{j},\nn\\
    A^{i}_{2}&=& \omega^{i}\omega^{j}\tilde{E}^{j},
	\label{veten}
\end{eqnarray}
Now, we calculate the required derivative of $\delta f$ and substitute them in the Eq.~\eqref{AA1} to solve for $C_{n}$ and $A_{n}$,
\begin{eqnarray}
	\frac{\partial \delta f}{\partial p^i} &=& \frac{\partial}{\partial p_i}\sum_{n=0}^{6}C_nC_{n}^{kl} p^k p^l+\sum\limits_{n=0}^{2}A_{n}~A_{n}^{k}~p^{k}\nn\\
	&=& \sum_{n=0}^{6}C^{kl}_{n}p^k p^l\frac{\partial C_n}{\partial p^i}+ \sum_{n=0}^{6} C^{kl}_{n} C_n \frac{\partial }{\partial p^i} p^k p^l+\sum_{n=0}^{2}A^{k}_{n}p^k \frac{\partial A_n}{\partial p^i}+ \sum_{n=0}^{2} A^{k}_{n} A_n \frac{\partial }{\partial p^i} p^k .\nn
\end{eqnarray}
It can easily be seen that the $C_n$ and $A_{n}$ for which Eq.~\eqref{AA1} is satisfied are functions of $f^0$, i.e., $C_{n}=C_n(f^0)$ and $A_{n}(f^0)$.
\begin{equation}
	\frac{\partial \delta f}{\partial p^i} =-\frac{1}{T}f^0(1+\xi f^0)\sum_{n=0}^{6} \frac{dC_n}{df^0} C^{kl}_{n} \frac{p^k p^l p^i}{E} +\sum_{n=0}^{6}2C_nC^{ik}_{n}p^{k}-\frac{1}{T}f^0(1+\xi f^0)\sum_{n=0}^{2} \frac{dA_n}{df^0} A^{k}_{n} \frac{p^k p^i}{E} +\sum_{n=0}^{2}A_nA^{i}_{n}
	\label{AA3}
\end{equation}

Using the result of Eq.~\eqref{AA3} in Eq.~\eqref{AA1} we have:
\begin{eqnarray}
	&&\left[\frac{ p^i p^j}{ET} U^{ij} f^0(1+\xi f^0)-\frac{1}{ \tau_{\Om}T}f^0(1+\xi f^0)\sum_{n=0}^{6}  \frac{dC_n}{df^0} \om^{ij} \frac{p^{i}p^{j}p^{k}p^{l}}{E}C_{n}^{kl} + \frac{2}{\tau_\Om} \sum_{n=0}^{6}  C_n C^{ik}_n \om^{ij}p^{j}p^{k}\right]\nn\\
	&+&\left[-\frac{q\tilde{E}^{i}p^{i}}{ET}f^0(1+\xi f^0)-\frac{1}{\tau_{\Omega}T}f^0(1+\xi f^0)\sum_{n=0}^{2} \frac{dA_n}{df^0}  \omega^{ij}\frac{p^jp^k p^i}{E} A^{k}_{n} +\frac{1}{\tau_{\Omega}}\sum_{n=0}^{2}A_nA^{i}_{n}\omega^{ij}p^{j}\right]=\left[-\frac{1}{\tau_{c}}\sum_{n=0}^{6} C_{n} C^{kl}_{n} p^{k} p^{l}\right]\nn\\
	&+&\left[-\frac{1}{\tau_{c}}\sum\limits_{n=0}^{2}A_{n}~A_{n}^{k}~p^{k}\right]\nn\\	   
	\implies && \left[\frac{ p^i p^j}{ET} U^{ij} f^0(1+\xi f^0) + \frac{2}{\tau_\Om} \sum_{n=0}^{6}  C_n C^{ik}_n \om^{ij}p^{j}p^{k}\right] +\left[-\frac{q\tilde{E}^{i}p^{i}}{ET}f^0(1+\xi f^0)+\frac{1}{\tau_{\Omega}}\sum_{n=0}^{2}A_nA^{i}_{n}\omega^{ij}p^{j}\right]\nn\\
	&=&\left[-\frac{1}{\tau_{c}}\sum_{n=0}^{6} C_{n} C^{kl}_{n} p^{k} p^{l}\right]+\left[-\frac{1}{\tau_{c}}\sum\limits_{n=0}^{2}A_{n}~A_{n}^{k}~p^{k}\right], (\text{ since, }\om^{ij}p^{i}p^{j}=0)~.\label{BTEtotal}    
\end{eqnarray}	
From Eq.~\eqref{BTEtotal} we get two independent equation to solve for the viscosity and conductivity sectors they are given by,
\begin{eqnarray}
&& \frac{ p^i p^j}{ET} U^{ij} f^0(1+\xi f^0)  = \sum_{n=0}^{6} C_n \left(-\frac{2} {\tau_\Om} C^{ik}_{n}\om^{ij} p^j p^k -\frac{1}{\tau_c} C^{kl}_{n} p^{k}p^{l}\right),
	\label{AA4}\\
	&& -\frac{q\tilde{E}^{i}p^{i}}{ET}f^0(1+\xi f^0)=\sum_{n=0}^{2}A_n\left(-\frac{1}{\tau_{\Omega}}A^{i}_{n}\omega^{ij}p^{j}-\frac{1}{\tau_{c}}~A_{n}^{k}~p^{k}\right)~.\label{sigmaeq}
\end{eqnarray}
The Eq.~\eqref{AA4} has to be solved for the $C_{n}$ to obtain $\delta f$ for the evaluation of viscosities. In the present article our aim is to obtain shear viscosities of the system therefore we will ignore $C_{5}$ and $C_{6}$ which correspond to bulk viscosities and retain the first five $C_n$ that correspond to shear stresses in the fluid. By equating the coefficients of $p^{i}p^{j}U^{ij} ,U^{ij}p^{j}p^{k}\om^{ik},U^{ij}p^{k}\om^{j}\om^{ik}(\vec{p}\cdot\vec{\om})\text{ and } U^{ij}p^i\om^j(\vec{p}\cdot\vec{\om})$ in Eq.~\eqref{AA4} to zero, we have,
\begin{eqnarray}
p^{i}p^{j}U^{ij} &:& -\frac{4C_3}{\tau_{\Om}} -\frac{2C_1}{\tau_c} = \frac{1}{ET}f^0(1+\xi f^0)~,\nn\\
U^{ij}p^{j}p^{k}\om^{ik} &:& -\frac{4C_1}{\tau_{\Om}}+\frac{2C_3}{\tau_c}=0~,\nn\\
U^{ij}p^{k}\om^{j}\om^{ik}(\vec{p}\cdot\vec{\om})&:& \frac{4C_1}{\tau_{\Om}}-\frac{4C_2}{\tau_\Om}-\frac{2C_3}{\tau_{c}}+\frac{4C_4}{\tau_c}=0~,\nn\\
U^{ij}p^i\om^j(\vec{p}\cdot\vec{\om}) &:& \frac{8C_3}{\tau_{\Om}}-\frac{4C_4}{\tau_\Om}+\frac{4C_1}{\tau_{c}}-\frac{4C_2}{\tau_c}=0~.
\label{AA5}
\end{eqnarray}
Solving the above set of linear equations we obtain,
\begin{eqnarray}
	&&C_1=-\frac{1}{2ET}f^0(1+\xi f^0)\frac{\tau_c}{1+4(\tau_c/\tau_\Om)^2}~,\nn\\
	&&C_2=-\frac{1}{2ET}f^0(1+\xi f^0)\frac{\tau_c}{1+(\tau_c/\tau_\Om)^2}~,\nn\\
	&&C_3=-\frac{1}{ET}f^0(1+\xi f^0)\frac{\tau_c(\tau_c/\tau_\Om)}{1+4(\tau_c/\tau_\Om)^2}~,\nn\\
	&&C_4=-\frac{1}{2ET}f^0(1+\xi f^0)\frac{\tau_c(\tau_c/\tau_\Om)}{1+4(\tau_c/\tau_\Om)^2}~.
	\label{AA6}
\end{eqnarray}
Similarly solving Eq.~\eqref{sigmaeq} by equating the coefficients of $\om^{i}\om^j\tilde{E}^{j}p^{i}$, $\omega^{ij}\tilde{E}^{j}p^{i}$ and $\tilde{E}^{i}p^{i}$ to zero, we have,
\begin{eqnarray}
	\omega^{ij}\tilde{E}^{j}p^{i} &:& \frac{A_0}{\tau_{\Om}}-\frac{A_1}{\tau_c}=0~,\nn\\
	\tilde{E}^{i}p^{i}&:& \frac{A_1}{\tau_{\Om}}+\frac{A_0}{\tau_c}=\frac{q}{ET}f^0(1+\xi f^0)~, \nn\\
	\om^{i}\om^j\tilde{E}^{j}p^{i} &:& \frac{A_1}{\tau_{\Om}} -\frac{A_2}{\tau_c}=0~,
	\label{sigmasolu}
\end{eqnarray}
Simplifying the above set of linear equations we obtain, 
\begin{eqnarray}
	&&A_n=\frac{q}{ET}\frac{\tau_c\left(\frac{\tau_{c}}{\tau_{\Omega}}\right)^{n}}{1+\Big(\frac{\tau_c}{\tau_{\Omega}}\Big)^2} f^0(1+\xi f^0)~.
	\label{sigmasolex}
\end{eqnarray}

\section{Calculation of shear viscosity and entropy density for massless QGP}\label{ape1}
\begin{eqnarray}
	\eta^{\rm QGP}_{0}\equiv \eta ^{\rm QGP}&=&\frac{g_{q}\tau_{c}}{15T}\int \frac{d^{3}p}{(2\pi)^{2}} \frac{p^{4}}{E^{2}}f^{0}_{q}(1-f^{0}_{q})	+\frac{g_{g}\tau_{c}}{15T} \int \frac{d^{3}p}{(2\pi)^{2}} \frac{p^{4}}{E^{2}}f^{0}_{g}(1+f^{0}_{g})\nn\\
	&=&\frac{g_{q}\tau_{c}}{30\pi^{2}T}\int dE~ E^{4}f^{0}_{q}(1-f^{0}_{q})+\frac{g_{g}\tau_{c}}{30\pi^{2}T} \int  dE~ E^{4}f^{0}_{g}(1+f^{0}_{g})\nn\\
	&=&\frac{\tau_{c}T^{4}}{30\pi^{2}} 24T\frac{\del}{\del \mu} (g_{q}f^{\rm FD}_{5}(A)+g_{g} f^{\rm BE}_{5}(A)), \big(\text{ where, } f^{\rm FD/BE}_j(A)=\frac{1}{\Gamma(j)}\int_{0}^{\infty}\frac{x^{j-1}dx}{A^{-1}e^{x}\pm 1}, (A\equiv e^{\mu/T})\big)\nn\\
	&=& \frac{4\tau_{c}T^{4}}{5\pi^{2}} \left(g_{q}f^{\rm FD}_{4}(1)+g_{g}f^{\rm BE}_{4}(1)\right)\quad \quad (\text{by assuming }\mu=0,A=1 )\nn\\
	&=& \frac{4\tau_{c}T^{4}}{5\pi^{2}} \left(g_{q}\left(1-\frac{1}{2^{4-1}}\right) + g_{g}\right)\zeta(4)\nn\\
	&=& \frac{4\tau_{c}T^{4}}{5\pi^{2}}\left[\frac{7}{8}g_{q}+g_{g}\right]\zeta(4)=\frac{19 \pi^{2}}{45}\tau_{c}T^{4}\label{isoQap}
\end{eqnarray}
where we used $g_{q}=3 (\rm flavor) \times 3 (\rm color) \times 2(\rm spin) \times 2 (\rm particle-antiparticle)=36$, $g_{g}=2 (\rm polarization) \times 8 (\rm color)=16$ and the $\zeta(4)=\frac{\pi^{4}}{90}$. The pressure of the QGP can be evaluated as,
\begin{eqnarray}
	P^{\rm QGP}&=&g_{q}\int \frac{d^{3}p}{(2\pi)^{3}} \frac{p^{2}}{3E} f^{0}_{q}+g_{g}\int \frac{d^{3}p}{(2\pi)^{3}} \frac{p^{2}}{3E} f^{0}_{g}\nn\\
	&=&\frac{1}{6\pi^{2}} \left[g_{q}\int dE E^{3} f^{0}_{q}+ g_{g}\int dE E^{3} f^{0}_{g}\right]\nn\\
	&=&\frac{T^{4}}{\pi^{2}}\left[g_{q}\left(1-\frac{1}{2^{3}}\right)+ g_{g}\right]\zeta(4)=\frac{19\pi^{2}}{36}T^{4}~.\label{Qpr}
\end{eqnarray}
Using the fact that $\mathcal{E}^{\rm QGP}=3P^{\rm QGP}$ we have,
\begin{eqnarray}
	&&s^{\rm QGP}=\frac{\mathcal{E}^{\rm QGP}+P^{\rm QGP}}{T}=\frac{19\pi^{2}}{9}T^{3}.\label{entQ}
\end{eqnarray}
\end{widetext}

\section{Calibration of the relaxation time in QGP--HRG framework}\label{appeta_cal}
\begin{figure}[h!]
	\centering
	\includegraphics[width=\linewidth]{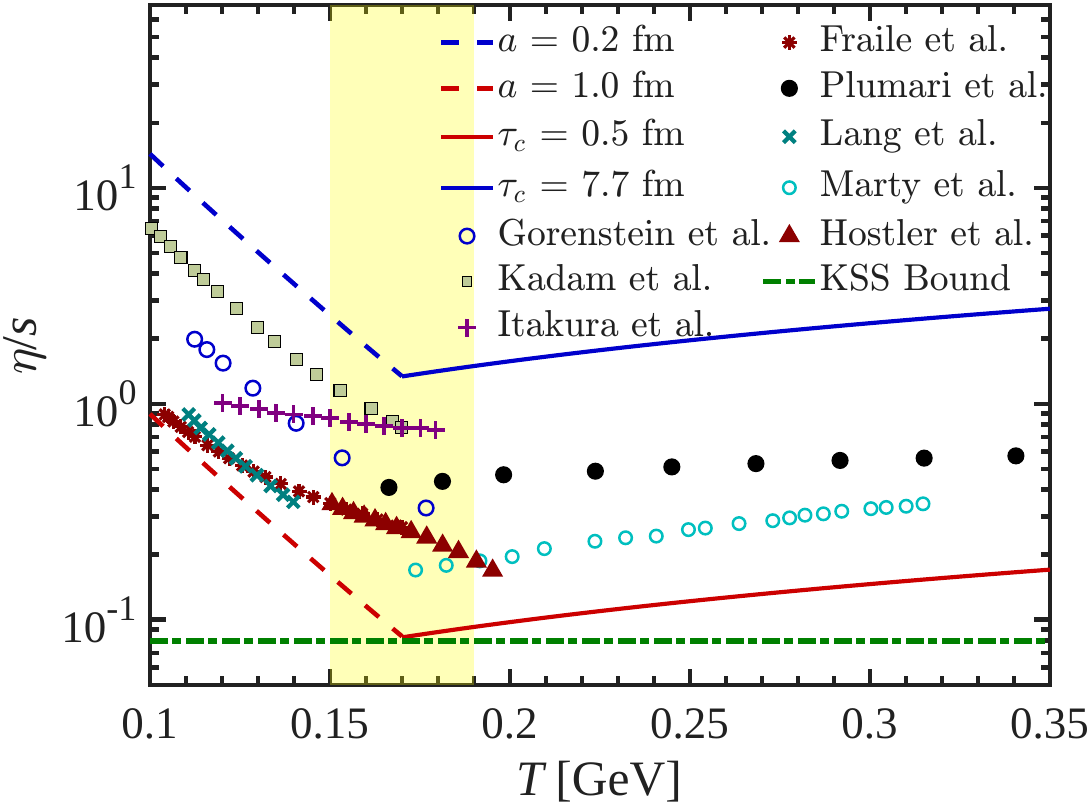}
	\caption{(Color online) The variation of shear viscosity to entropy density ratio ($\eta/s$) with temperature comparing results from different previous model calculations (Gorenstein \textit{et al.} ~\cite{Gorenstein:2007mw}, Kadam \textit{et al.}~\cite{Kadam:2014cua}, Itakura \textit{et al.} ~\cite{Itakura:2007mx}, Fraile \textit{et al.} ~\cite{Fernandez-Fraile:2009eug}, Plumari  \textit{et al.} \cite{Plumari:2012ep}, Lang \textit{et al.}~\cite{Lang:2012tt}, Marty \textit{et al.} ~\cite{Marty:2013ita}, Hostler \textit{et al.}  \cite{Noronha-Hostler:2008kkf}) along with the KSS bound obtained from AdS/CFT correspondence. The critical temperature is taken to be 0.17~GeV.}
	\label{Figure_1}
\end{figure}

Fig.~(\ref{Figure_1}) illustrates the temperature dependence of shear viscosity-to-entropy density ratio (\(\eta/s\)) for both the hadronic and QGP phases using the relaxation time approximated kinetic theory framework. Several previous microscopic estimates \cite{Gorenstein:2007mw, Kadam:2014cua,Itakura:2007mx,Fernandez-Fraile:2009eug,Plumari:2012ep,Lang:2012tt,Marty:2013ita,Noronha-Hostler:2008kkf} of shear viscosity-to-entropy density ratio (\(\eta/s\)) can be found in the literature employing various effective and quasiparticle theories of QCD matter. Results obtained in Gorenstein \textit{et al.} (VDW-HRG)~\cite{Gorenstein:2007mw}, Kadam \textit{et al.} (HRG-HS)~\cite{Kadam:2014cua}, Itakura \textit{et al.} (pion gas)~\cite{Itakura:2007mx}, Fraile \textit{et al.} (ChPT)~\cite{Fernandez-Fraile:2009eug}, Plumari  \textit{et al.} (gluon plasma-RTA) \cite{Plumari:2012ep}, Lang \textit{et al.} (pion gas)~\cite{Lang:2012tt}, Marty \textit{et al.} (DQPM)~\cite{Marty:2013ita}, Hostler \textit{et al.} (VDW-HRG) \cite{Noronha-Hostler:2008kkf} are shown in the figure. The theoretical lower limit on the shear viscosity-to-entropy density ratio -- the KSS bound (\(\eta/s \geq 1/4\pi\)) -- obtained from AdS/CFT correspondence~\cite{Kovtun:2004de} is also indicated for reference. To calibrate our estimations to behave similar to these model calculations we take the following model for the relaxation time.
 The relaxation time \(\tau_c\) is fixed at a temperature-independent constant value in the QGP phase, whereas in the hadronic phase, the relaxation time, which depends on both temperature and scattering length, is determined by,
\begin{eqnarray}
	&&\tau^{B,M}_{c}=\frac{1}{n_{\rm HRG}~v^{B,M}_{\rm av}~\pi a^{2}}\label{tau}~,
\end{eqnarray}
where $v^{B,M}_{\rm av}$ for any hadron is given by,
\begin{eqnarray}
	v^{B,M}_{\rm av}=\frac{\int \frac{d^3p}{(2\pi)^3}\frac{p}{E_{B,M}} f^{0}_{B,M}}{\int \frac{d^3p}{(2\pi)^3} f^{0}_{B,M}}~,\label{av_vel}
\end{eqnarray}
and the total number density for the HRG is expressed as,
\begin{eqnarray}
	&& n_{\rm HRG}=  \sum_{B} g_B\int \frac{d^{3}p}{(2\pi)^3}f_{B}^0
	+ \sum_{M} g_M\int \frac{d^{3}p}{(2\pi)^3}f_{M}^0~.\nn\\\label{NHRG}
\end{eqnarray}
  The maximum and minimum values of \(\tau_c\) in the QGP phase and scattering length \(a\) in the hadronic phase are tuned such that they are analytically continuous at $T_c$. At chemical potential $\mu = 0$, the phase transition in QCD is an analytic crossover~\cite{PhysRevD.85.054503}, so $\eta/s$ is expected to be continuous and smooth. The shaded region around the quark-hadron phase transition temperature $T_c\approx170$ MeV is used to remind the reader that the motivation of the present work is to provide a broad illustration of the shear viscosity across both phases. Our analysis mainly focuses on the order of magnitude and general trends of shear viscosity components outside this shaded region. This is because the region around the transition temperature can involve several uncertainties, such as (1) the precise location of the transition temperature—especially when using two different models for the two phases—and (2) how the transition temperature may be affected by rotation.  

To encompass a range of earlier $\eta/s$ estimations obtained without considering rotation, from various effective transport models, we span the scattering length from $0.25$ fm to $1$ fm in the hadronic phase, while the relaxation time is varied from $0.48$ to $7.75$ fm in the QGP phase, as shown in Fig.~(\ref{Figure_1}). The red (blue) solid curves give the lower (upper) limit from our relaxation time approximated kinetic theory estimates in the HRG domain, and the red (blue) dot-dashed curves give the lower (upper) limit in the QGP phase. The behavior of \(\eta/s\) reveals a decreasing trend in the hadronic phase, while in the QGP phase, they increase monotonically.

\section{Calibration of the relaxation time in the NJL framework}\label{appeta_cal1}
Similar to the calibration performed in the appendix~\eqref{appeta_cal}, we model the relaxation time in the NJL model to cover the earlier estimations of shear viscosity and electrical conductivity. 
For the temperature $T> 170$ MeV, where one expects deconfined quark matter, we assume $\tau_{c}$ to be a tunable constant. For the temperature $T<170$ MeV, we calculate $\tau_{c}$ using a hard-sphere scattering relation,  
\begin{equation}
	\tau_{c}=\frac{1}{n~\pi a^{2}~v_{\text{av}}}~,\label{tauh}
\end{equation}
where $a$ is the scattering length treated here as a parameter. The $n$ and $v_{\text{av}}$ are, respectively, the total quark density and thermal average velocity of the quarks given by,
\begin{eqnarray}
	&& n = 24\int{\frac{d^{3}\vec{p}}{{(2\pi})^{3}}} ~f^{0},\nn\\
	&& v_{\text{av}} = \frac{\int{\frac{d^{3}\vec{p}}{{(2\pi})^{3}}} \frac{p}{E}~f^0}{\int{\frac{d^{3}\vec{p}}{{(2\pi})^{3}}} ~f^{0}},
\end{eqnarray}
where $f^{0}=1/[e^{E/T}+1]$ is the Fermi-Dirac distribution for the constituent quarks with energy $E=\sqrt{\vec{p}^{2}+M^{2}}$.
The realistic range for the $\tau_{c}$ in the whole temperature range is found by covering the numerical magnitudes of the $\sigma/T$ and $\eta/s$ obtained by different authors~\cite{Cassing:2013iz,Fernandez-Fraile:2009eug,Greif2014,Marty:2013ita,Puglisi:2014pda,Amato2013,Gorenstein:2007mw,Itakura:2007mx,Plumari:2012ep, Noronha-Hostler:2008kkf} at $\Om=0$. We used the results of the following papers, Cassing \textit{et al.} (PHSD)~\cite{Cassing:2013iz}, Fraile \textit{et al.} (ChPT)~\cite{Fernandez-Fraile:2009eug}, Greif \textit{et al.} (BAMPS)~\cite{Greif2014}, Marty \textit{et al.} (NJL)~\cite{Marty:2013ita}, Marty \textit{et al.} (DQPM)~\cite{Marty:2013ita}, Puglisi \textit{et al.} (PQCD)~\cite{Puglisi:2014pda}, Puglisi \textit{et al.} (QP)~\cite{Puglisi:2014pda}, Amato \textit{et al.} (LQCD)~\cite{Amato2013} to calibrate the relaxation time for conductivity.
\begin{figure}[H]
	\centering
	\includegraphics[width=\linewidth]{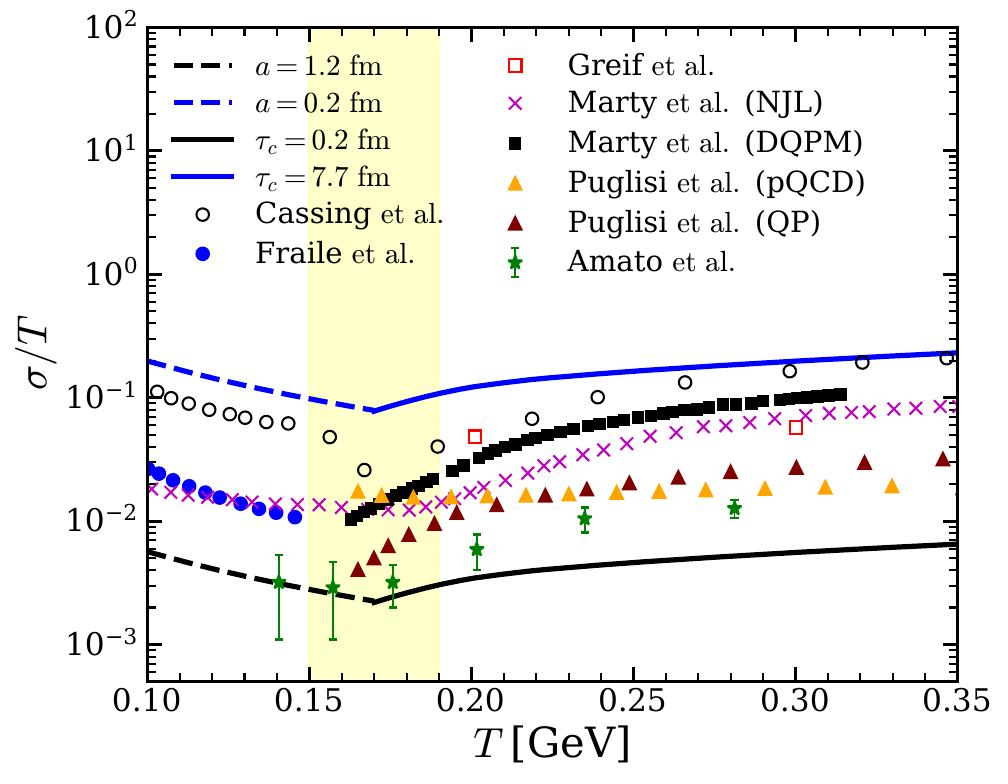}
	\caption{Calibration of $\tau_{c}$ from comparison with existing calculation of conductivity at $\Omega=0$.}
	\label{sigmatau}
\end{figure} 
 Similarly, we used the results of the following papers:  Itakura \textit{et al.} (pion gas)~\cite{Itakura:2007mx}, Fraile \textit{et al.} (ChPT) ~\cite{Fernandez-Fraile:2009eug}, Plumari \textit{et al.} (gluon plasma-RTA) ~\cite{Plumari:2012ep}, Marty \textit{et al.} (DQPM)~\cite{Marty:2013ita}, Gorenstein \textit{et al.} (VDW-HRG)~\cite{Gorenstein:2007mw}, Hostler \textit{et al.} (VDW-HRG)~\cite{Noronha-Hostler:2008kkf} to calibrate the relaxation time for shear viscosity. 
\begin{figure}[H]
	\centering
	\includegraphics[width=\linewidth]{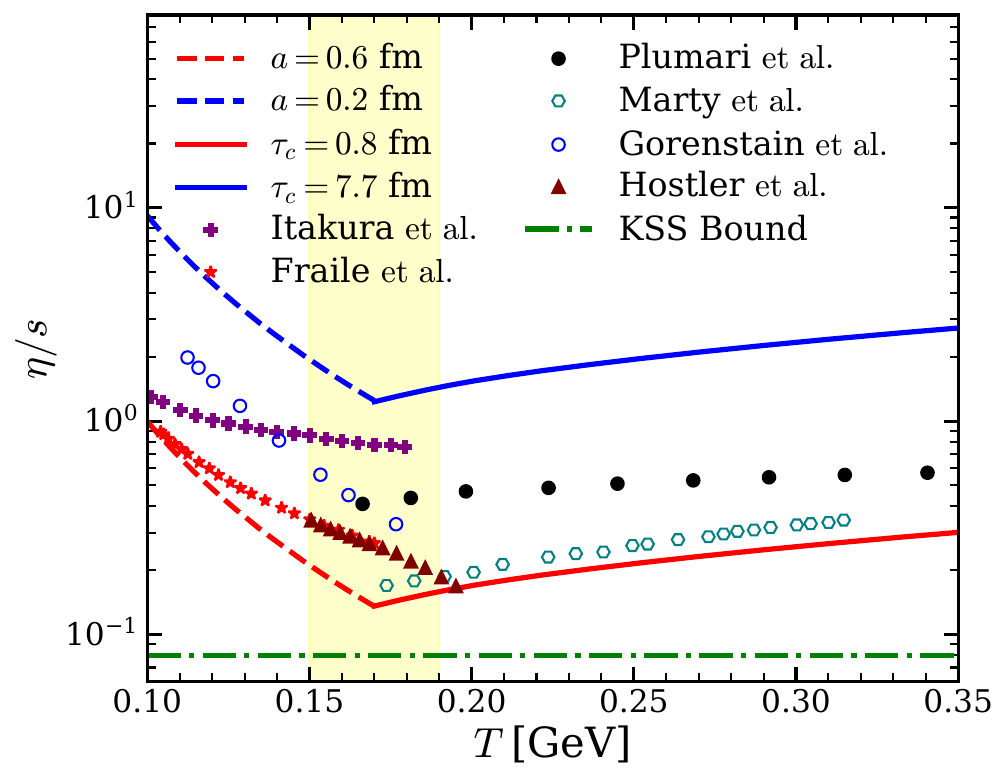}
	\caption{Calibration of $\tau_{c}$ from comparison with existing calculation of viscosity at $\Omega=0$.}
	\label{etatau}
\end{figure}
In Fig.~(\ref{sigmatau}), we have calibrated the scattering length $a$ for the temperature domain $T<170$ MeV to cover the numerical band of the $\sigma/T$ obtained in the Refs.~\cite {Cassing:2013iz,Fernandez-Fraile:2009eug,Marty:2013ita}. Similarly, for the temperature range $T>170$ MeV, we tuned the relaxation time $\tau_{c}$ to cover the numerical values of $\sigma$. It is evident from Fig.~(\ref{sigmatau}) that for $\tau_{c}\in (0.2,7.7)$ fm and $a\in(0.2,1.2)$ fm, the NJL model estimations cover the results of many authors in the whole temperature range. The same calibration procedure for $\eta/s$ in Fig.~(\ref{etatau}) results in $\tau_{c}\in(0.8,7.7)$ fm and $a\in (0.2, 0.6)$ fm. The shaded yellow band in Fig.~(\ref{sigmatau}) and 
Fig.~(\ref{etatau}) around $T=170$ MeV is to remind the readers that our main focus in the present paper is to see the overall trend of shear viscosity and electrical conductivity in the high-temperature (chirally-symmetry restored) and low-temperature (chirally-symmetry broken) regime. Although we have smoothly matched the curves of two phases at T = 170 MeV, the analysis mainly focuses on the order of magnitude and qualitative behavior of shear viscosity and electrical conductivity components outside this shaded region. We end the discussion with the note that for $T<T_{c}$ the thermodynamic phase space part of the transport coefficients in NJL and HRG are not same. In NJL model fermionic degrees of freedom (quarks) dominate the phase space part whereas in HRG mostly the lighter bosonic (pions and kaons) degrees of freedom contributes. Nevertheless, transport coefficients depend not only on this phase space part but also on the relaxation time. Therefore; one may get a similar estimate for the transport coefficients by choosing dissimilar scattering lengths or equivalently relaxation time in NJL and HRG.
\bibliographystyle{unsrturl}
\bibliography{HRG_NJL_combined}

\end{document}